\documentclass[a4paper]{aa}

\usepackage{rotating}
\usepackage{acronym}
\usepackage{xspace}
\usepackage{multirow}
\usepackage{mathtools}
\usepackage{amsmath}
\usepackage{hyperref}
\usepackage{units}
\usepackage{threeparttablex}
\usepackage{natbib,twoopt}

\usepackage{color}
\usepackage{graphicx}
\usepackage{amsmath}
\usepackage{amssymb}
\usepackage{appendix}
\usepackage{longtable}
\usepackage{lineno}

\definecolor{chmagenta}{rgb}{0.54, 0.17, 0.88}

\graphicspath{{./}{figures/}}

\begin{document}

\title{The Host Galaxies of High Velocity Type Ia Supernovae}


\author{{Anya E. Nugent}
          \inst{1}
          \and
          {Abigail E. Polin}\inst{2,3}
          \and
          {Peter E. Nugent} \inst{4}
          }

   \institute{{Center for Interdisciplinary Exploration and Research in Astrophysics (CIERA) and Department of Physics and Astronomy, Northwestern University, Evanston, IL 60208, USA\\
              \email{anyanugent2023@u.northwestern.edu}}
              \and
            Carnegie Observatories, 813 Santa Barbara Street, Pasadena, CA 91101, USA
            \and
            TAPIR, Walter Burke Institute for Theoretical Physics, 350-17, Caltech, Pasadena, CA 91125, USA
            \and
            Lawrence Berkeley National Lab, 1 Cyclotron Road, Berkeley, CA 94720, USA
              }
            
\abstract{
In recent years, there has been ample evidence that Type Ia supernova (SNe~Ia) with high \ion{Si}{2} velocities near peak brightness are distinguished from SNe~Ia of lower velocities and may indeed represent a separate progenitor system. These SNe~Ia can contaminate the population of normal events used for cosmological analyses, creating unwanted biases in the final analyses. Given that many current and future surveys using SNe~Ia as cosmological probes will not have the resources to take a spectrum of all the events, likely only getting host redshifts long after the SNe~Ia have faded, we need to turn to methods that could separate these populations based purely on photometry or host properties. Here, we present a study of a sample of well observed, nearby SNe~Ia and their hosts to determine if there are significant enough differences between these populations that can be discerned only from the stellar population properties of their hosts. Our results indicate that the global host properties, including star formation, stellar mass, stellar population age, and dust attenuation, of high velocity SNe~Ia do not differ significantly from those of lower velocities. However, we do find that high velocity SNe~Ia are more concentrated toward the center of their hosts, suggesting that their local environments may indeed differ. Future work requires strengthening photometric probes of high velocity SNe~Ia and their local environments to distinguish these events and determine if they originate from a separate progenitor.}

\keywords{type Ia supernovae, galaxies, white dwarfs}

\maketitle
\section{Introduction}
\label{sec:intro}
Type Ia supernovae (SNe~Ia) are well-characterized as standardized candles given how their their peak luminosities, colors, and light curves relate to distance, and have thus been powerful tools in determining the expansion rate of the Universe \citep{1998AJ....116.1009R,1999ApJ...517..565P}. However, the reasoning behind the variations in their light curves and luminosities, which affects our precision on cosmology, remains largely unknown, posing a dilemma in determining how best to and which SNe~Ia can be standardized. Currently, all cosmology measurements using SNe~Ia are limited by systematic, not statistical, uncertainties \citep{sds2020}.

One current predicament in SNe~Ia cosmology is understanding if SNe with high \ion{Si}{2} velocities at $\sim$6150\AA\ ($\gtrsim 12,000$~km~s$^{-1}$, measured at the time of peak brightness) can be used alongside lower velocity SNe in cosmology measurements. Several key differences between high and lower velocity SNe~Ia have been discussed extensively in the literature. Notably, the high velocity SNe~Ia sample appears have a much narrower distribution of expected $M_B$ versus \ion{Si}{2} velocities than the lower velocity SNe sample and redder $B-V$ colors \citep{2009ApJ...699L.139W, fk2011, pnk2019}. Recent work has suggested that multiple progenitor systems may account for differences in observed luminosities and light curves \citep{bulla20} of SNe~Ia, which would complicate the systematics in their use as a precision tool for cosmology. For example, it has often been proposed that SNe~Ia derive from the thermonuclear explosion of a white dwarf that approaches the Chandrasekhar mass limit (see \citealt{maoz2014} for a review). However, there exists both observational and theoretical evidence that many SNe~Ia derive from multiple progenitor channels. These include hydrogen-rich SNe~Ia-CSM \citep{dhc+2012, hnh+2018}, super-Chandrasekhar \citep{2006Natur.443..308H,Hsaio2020} and sub-Chandrasekhar (sub-$M_\textrm{ch}$; \citealt{srs+14, gk2018, skm+2018, pnk2019, lmp+2023, nmd+2023}) mass explosions. Indeed, \citet{pnk2019} found that high velocity SNe~Ia likely derive from sub-$M_\textrm{ch}$ explosions, which naturally produce intrinsically redder SNe. However, alternate theories on the origins of high velocity SNe~Ia suggest that their host galaxies' global and local environmental properties are the true cause of their observed difference and simply need a different total to selective extinction ratio, $R_V$, to be properly standardized. \citet{2009ApJ...699L.139W}, for example, claims that their redder colors are likely due to local, dustier environments, and \citet{wang2013} showed they do indeed occur in denser and brighter regions of their hosts. In contrast, \citet{fk2011} proposes that the redder color is an intrinsic color difference. Since several current and many future surveys using SNe~Ia for cosmological measurements will be limited to purely photometric data \citep{2023MNRAS.518.1106V}, it is crucial to consider if the host galaxy properties of these SNe~Ia can be used to separate these sub-types and, more conclusively determine if the differences between these normal and high velocity populations are intrinsic to the progenitor or related to the local environment. 

Previous host galaxy analysis of SNe~Ia have been instrumental in both progenitor channel studies as well as providing corrections to the observed luminosities of these events for cosmology. The association of SNe~Ia with a diverse set of host environments, ranging from star forming to quiescent and low to high mass galaxies, has more conclusively secured their older stellar progenitor origins with a breadth of delay times. The ``mass-step" predicament, wherein SNe~Ia in higher mass galaxies ($\geq 10^{10}$M$_\odot$) are observed to be more luminous than those in lower mass galaxies, furthermore has lead to debates on whether properties of the host galaxy or the SNe~Ia progenitor affect the peak luminosity of such events \citep{kelly2010, sullivan2010, gupta2011, childress2013}. Several studies, for instance, claim that the dust relations in higher mass galaxies cause the observed difference and thus the extinction corrections to SNe~Ia in higher mass galaxies should also be different in order for these SNe to be properly standardized \citep{salim2018, bs2021, meldorf2023}. Relations between the width of the SNe light curve and host galaxy stellar population age, gas-phase metallicity, and local and total star formation rates (SFR), have also been observed, implying that age of the progenitor might be linked to SNe observables \citep{sullivan2006,neill2009,sullivan2010, lampeitl2010,dgs+2011,gupta2011, childress2013, pan2014,rigault2020}. Thus, a uniform host galaxy comparison between high and low velocity SNe~Ia is crucial to inform: (i) if the redness observed in high velocity SNe~Ia is due to the host environment or the progenitor; (ii) if the high velocity SN~Ia progenitor traces different stellar population properties (e.g. age, mass, star formation, etc.) than lower velocity SNe~Ia; and (iii) if global host galaxy properties can be used to separate these sub-classes.

Here, we model and determine the host galaxy stellar population properties of 74 Type Ia SNe, 14 of which are high velocity SNe~Ia, 56 of which are low velocity SNe~Ia, and 4 of which are SN 1991bg-like \citep{frb+1992}. By comparing the host samples and providing uniform stellar population modeling of these categories of SNe~Ia, we infer if the redder colors of high velocity SNe~Ia and their observable differences from low-velocity SNe~Ia can be explained by any host galaxy property. In Section \ref{sec:sample}, we describe the SNe~Ia sample and how they were selected for this study. We detail our archival photometry search for the hosts of the 74 SNe~Ia and the stellar population models used for this analysis in Section \ref{sec:SED}. We discuss our comparisons of host stellar population properties between the several SNe classifications in Section \ref{sec:results}. We discuss possible differences in the local environments of these SNe using their offsets from their host center and Na I~D equivalent widths in Section \ref{sec:local}. We comment on possible separate progenitor systems in Section \ref{sec:discussion}. Finally, our conclusions are in Section \ref{sec:conclusion}.

Unless otherwise stated, all observations are reported in the AB magnitude system and have been corrected for Galactic extinction in the direction of the SN \citep{MilkyWay,sf11}.  We employ a standard WMAP9 cosmology of $H_{0}$ = 69.6~km~s$^{-1}$~Mpc$^{-1}$, $\Omega_\textrm{M}$ = 0.286, $\Omega_\textrm{vac}$ = 0.714 \citep{Hinshaw2013, blw+14}. 

\begin{figure}[t]
\centering
\includegraphics[width=0.45\textwidth]{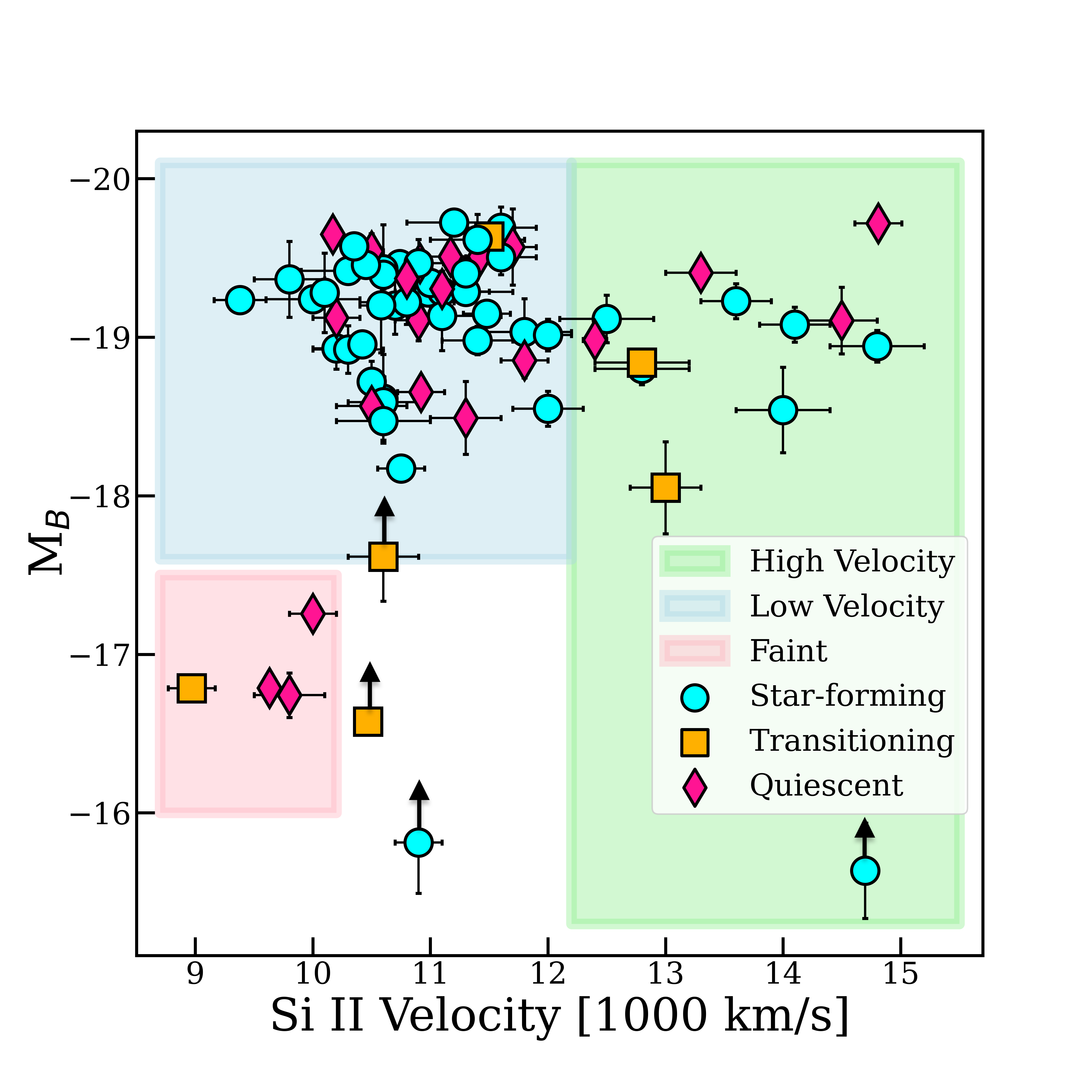}
\caption{The peak $B$-band absolute magnitudes ($M_B$) vs \ion{Si}{2} velocities with the uncertainties of 74 SNe from the \citet{zkf2017} and \citet{pnk2021} samples. The SNe are colored by their hosts' star formation classification: star-forming (blue circles), transitioning (yellow squares) and quiescent (red diamonds). High velocity SNe~Ia ($\geq 12,200$~km/s) are outlined in the green box, low velocity SNe~Ia in the blue box, and SN 1991bg-like faint SNe ($M_B > -17.5$~mag) are highlighted in the pink box. Arrows represent SNe~Ia that are not intrinsically faint, but rather fainter from some extinction. The entire population is dominated by star-forming galaxies ($64\%$), and the high and low velocity SNe~Ia roughly follow those fractions ($57\%$ and $70\%$, respectively).}
\label{sf_class}
\end{figure}

\begin{figure}[t]
\centering
\includegraphics[width=0.5\textwidth]{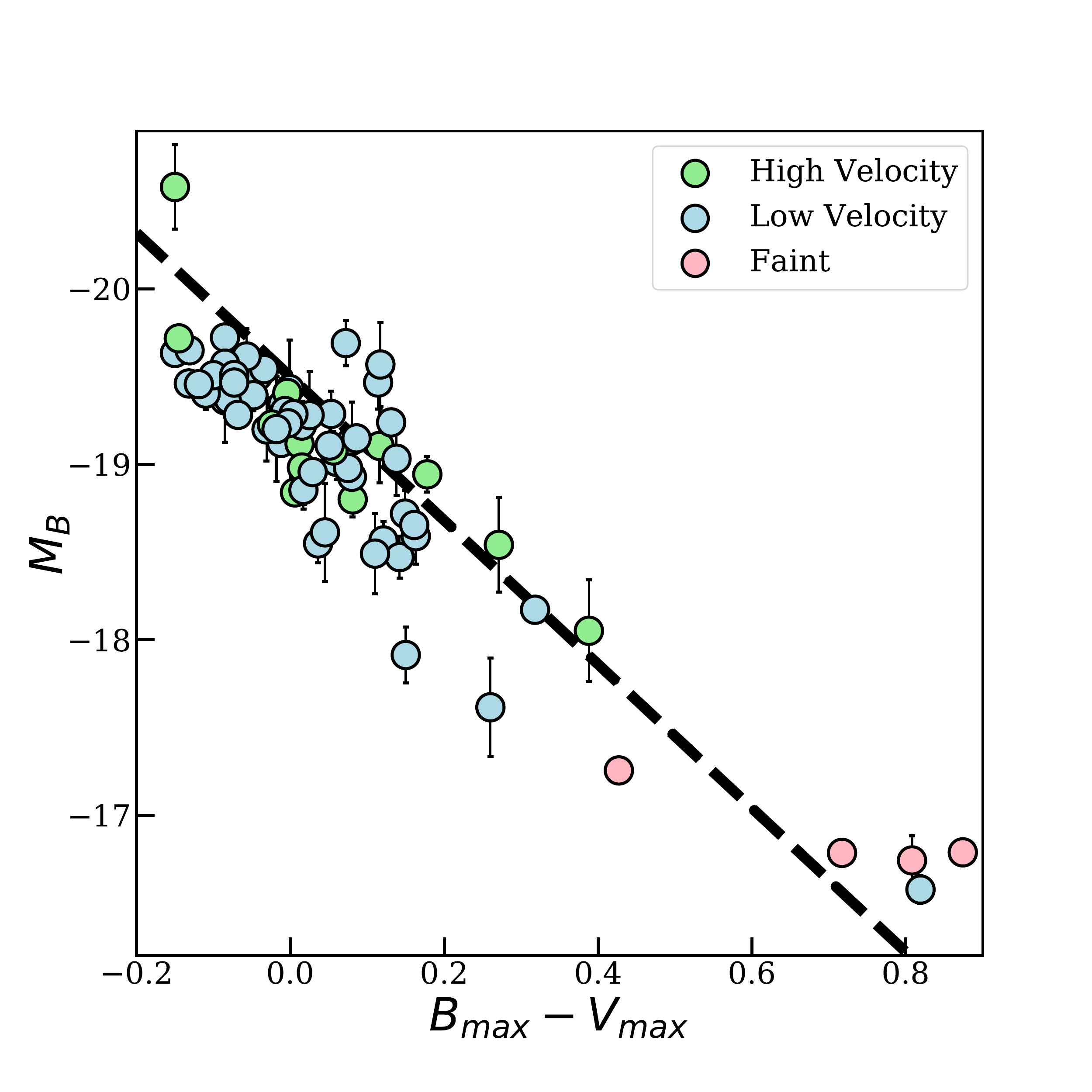}
\vspace{-0.3in}
\caption{The peak $B$-band absolute magnitudes and $B-V$ colors for high velocity (green), low velocity (blue), and faint SNe (pink). The dashed black line shows the standard Milky Way dust extinction law with $R_B=4.1$ and the $M_B$ intercept represents the $M_B$ that $B-V=0$ for unextinguished SNe~Ia in the SNooPy algorithm \citep{2011AJ....141...19B}. Given that most SNe~Ia are extinguished, we do expect that the majority of SNe~Ia in our sample will fall below this relation. High velocity SNe~Ia trend towards redder colors than low velocity SNe~Ia.}
\label{lum}
\end{figure}

\section{Supernovae Sample}
\label{sec:sample}
We conduct our host galaxy analysis based on the SNe~Ia sample described in \citet{zkf2017} and \citet{zkf2018}. This sample includes SNe discovered and observed by the Lick Observatory Supernovae Search (LOSS; \citealt{flt+2001, llcf2011}), the Harvard Smithsonian Center for Astrophysics Data Release 3 \citep{hcj+2009}, and the Carnegie Supernova Project \citep{chp+2010}. We note that these surveys are galaxy-targeted and can be constructed to be volume-limited. Their main advantage over other surveys is that they are able to detect faint SNe~Ia several magnitudes below peak brightness, and thus can find SNe~Ia that have a much smaller contrast with their hosts and/or suffer from local extinction. As the goal of this work is to better understand if faintness and redness in SNe~Ia is correlated with host galaxy properties, this sample is preferred over the Palomar Transient Factory (PTF), Zwicky Transient Facility (ZTF), and other magnitude-limited surveys, which are naturally biased against detecting faint SNe or those with low contrast over their host \citep{fsn+2017}. These SNe were selected based on the conditions outlined in \citet{zkf2017}, which required SN discovery at 1 magnitude fainter in $B$-band than peak brightness and good photometric coverage post-discovery to properly model their lightcurves and estimate their peak luminosities. We include 53 out of the 56 total SNe in this sample based on the available SNe and host galaxy data. The three SNe not used in our analysis did not have sufficient host galaxy photometry for stellar population modeling (see Section \ref{sec:obs}). We additionally include 21 more SNe in \citet{pnk2021} and references therein that have public data in the Weizmann Interactive Supernova Data Repository (WISeREP; \citealt{WISeREP}) and the Open Supernovae Catalog (OSC; \citealt{OSC}). These SNe~Ia have a peak spectrum for Si II velocity measurements, and peak $B$-band magnitudes and $B-V$ colors, thus have comparable data to the \citet{zkf2017} sample, as well as sufficient host galaxy observations for stellar population modeling. All SNe~Ia have spectroscopically confirmed redshifts that range between $0.0008 < z < 0.04$. In Table \ref{tab:sndata}, we present all SN properties used in this work. For each SN, the B-V values and Si II velocity measurements are obtained from \citet{zkf2018} and \citet{pnk2021}. We correct the $B$-band absolute magnitudes in \citet{zkf2018} and \citet{pnk2021} using the Hubble Flow distance moduli for each SN host acquired from the NASA/IPAC Extragalactic Database (NED). For the majority of SN sample, we use distance moduli from \citealt{mould2000} (``Virgo + GA + Shapely" on NED) and for several others we use the values cited in \citet{saha2001, macri2001,  Blakeslee2010, riess2022}. We choose these values for consistency across sample and because these distance moduli are corrected most for peculiar velocities.

We divide the SNe sample into several categories based on the extinction-corrected peak $B$-band absolute magnitudes ($M_B$) and the \ion{Si}{2} velocities determined in \citet{zkf2018}: high velocity SNe~Ia (\ion{Si}{2} velocities $> 12,200$~km~s$^{-1}$), low velocity SNe~Ia  ($< 12,200$~km~s$^{-1}$), and faint 1991bg-like \citep{frb+1992} SNe ($M_B > -17.5$~mag), which we highlight in Figure \ref{sf_class}. We classify three SNe that have $M_B > -17.5$~mag as low velocity rather than faint SNe~Ia, as they have positive Na I D equivalent widths at four times the uncertainty at minimum, in contrast to the other faint SNe~Ia, (see Section \ref{sec:NaID}, Table \ref{tab:sndata}). This likely implies that they are not intrinsically faint, but rather fainter due to some extinction. Given these sample divisions, we find that $19\%$ of the sample are high velocity SNe~Ia, $76\%$ are low velocity SNe~Ia, and $5\%$ are faint SNe. We note that only $\approx2$ SNe overlap between high and low velocity SNe~Ia samples given the uncertainties on the \ion{Si}{2} velocities. As is shown in Figures \ref{sf_class} and \ref{lum}, the high velocity SNe~Ia appear to have a much narrower distribution of $M_B$ and \ion{Si}{2} velocities than low velocity SNe~Ia, with lower $M_B$, and redder colors. However, we note that choice of these sub-classes is not simply made from random cuts in $M_B$ and velocity space, as model-independent, statistical methods for dividing the SNe~Ia sample has proven these sub-classes are clearly distinguished. For example, using a hierarchical cluster analysis, \citet{benetti2005} observed a clear population difference between these three categories of SNe using ratios of the \ion{Si}{2} lines. \citet{bba+2020} further found these groups of SNe are distinguished between multiple photometric and spectroscopic properties of the SNe using Gaussian mixture models. As there is significant evidence that these SN sub-classes are separate and robust, we focus on the SN host environments to determine if we can use these properties to separate the SN classes and if they point to different progenitor or environmental differences that affect the SN observables, as has been suggested previously \citep{2009ApJ...699L.139W, fk2011, wang2013, pnk2019, pan2020, pan2022}.

\section{Host Galaxy Observations \& SED Modeling}
\label{sec:SED}

\subsection{Archival Data Collection}
\label{sec:obs}
We determine host galaxies for each SN through querying nearby ($\lesssim 5\arcmin$) galaxies at approximately the same redshift as the SN via the NED. To model the host galaxies' stellar population properties, we require broadband photometric observations in at least three different photometric filters covering at least two wavelength ranges (UV, optical, IR, and mid-IR), ensuring that the  spectral energy distribution (SED) is well sampled. We collect archival photometric observations for all host galaxies via NED. We obtain UV, IR, and mid-IR observations through the {\it GALEX} \citep{Galex}, Two-Micron All Sky Survey (2MASS) \citep{2MASS}, and Wide-field Infrared Survey Explorer (WISE; \citealt{WISE}) surveys, respectively. Optical data is from the Sloan Digital Sky Survey (SDSS: $ugriz$; \citealt{SDSS2020}) survey and and \textit{UBVRI} data available on NED. To mitigate inconsistencies in the photometry selected for the study, we select photometry based on extended source models or wide aperture sizes that, in theory, will encapsulate the majority of the each host's flux. For {\it GALEX} observations, we use the Kron galaxy photometry if it is available or the elliptical aperture photometry. For SDSS data, we use the elliptical ``Model" photometry as this is optimal galaxies. For 2MASS and WISE data, we use the profile-fit photometry for extended sources or, if that is not available, the photometry with the widest aperture size (ensuring that the same aperture size is used for each survey). We do not use any photometric data that is contaminated by foreground sources. We correct all photometry for Galactic extinction in the directions of the SNe \citep{MilkyWay,sf11}. Finally, we require impose a 10\% error floor, forcing all uncertainties to be $\gtrsim 10\%$ the flux density value, to ensure that no photometric observation is overweighted in the stellar population modeling and to mitigate systematic errors that may arise with large aperture photometry. We show all host galaxy photometry in Table \ref{tab:phot}.

\begin{figure*}[t]
\centering{\includegraphics[width=0.50\textwidth]{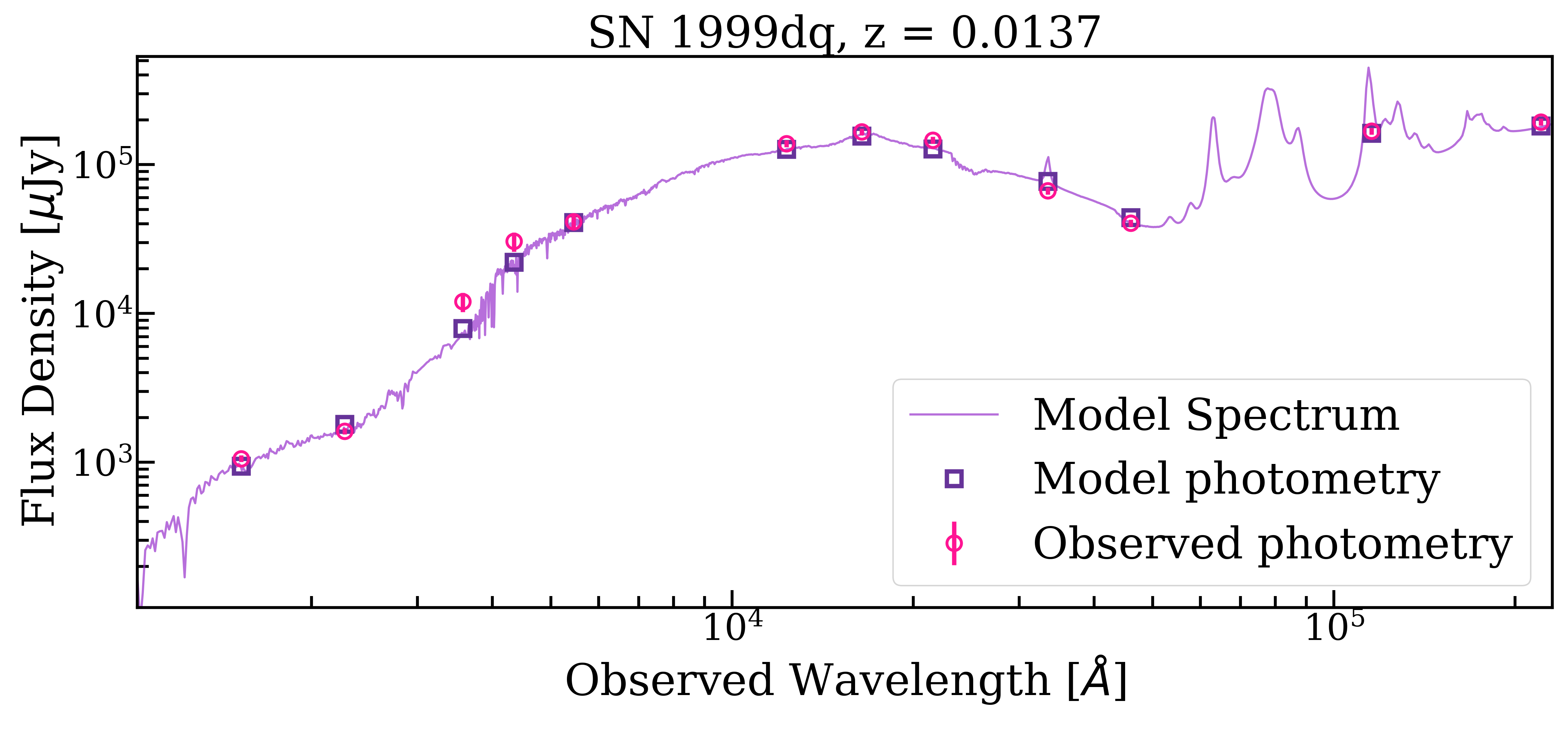}
\hspace{-0.1in}
\includegraphics[width=0.50\textwidth]{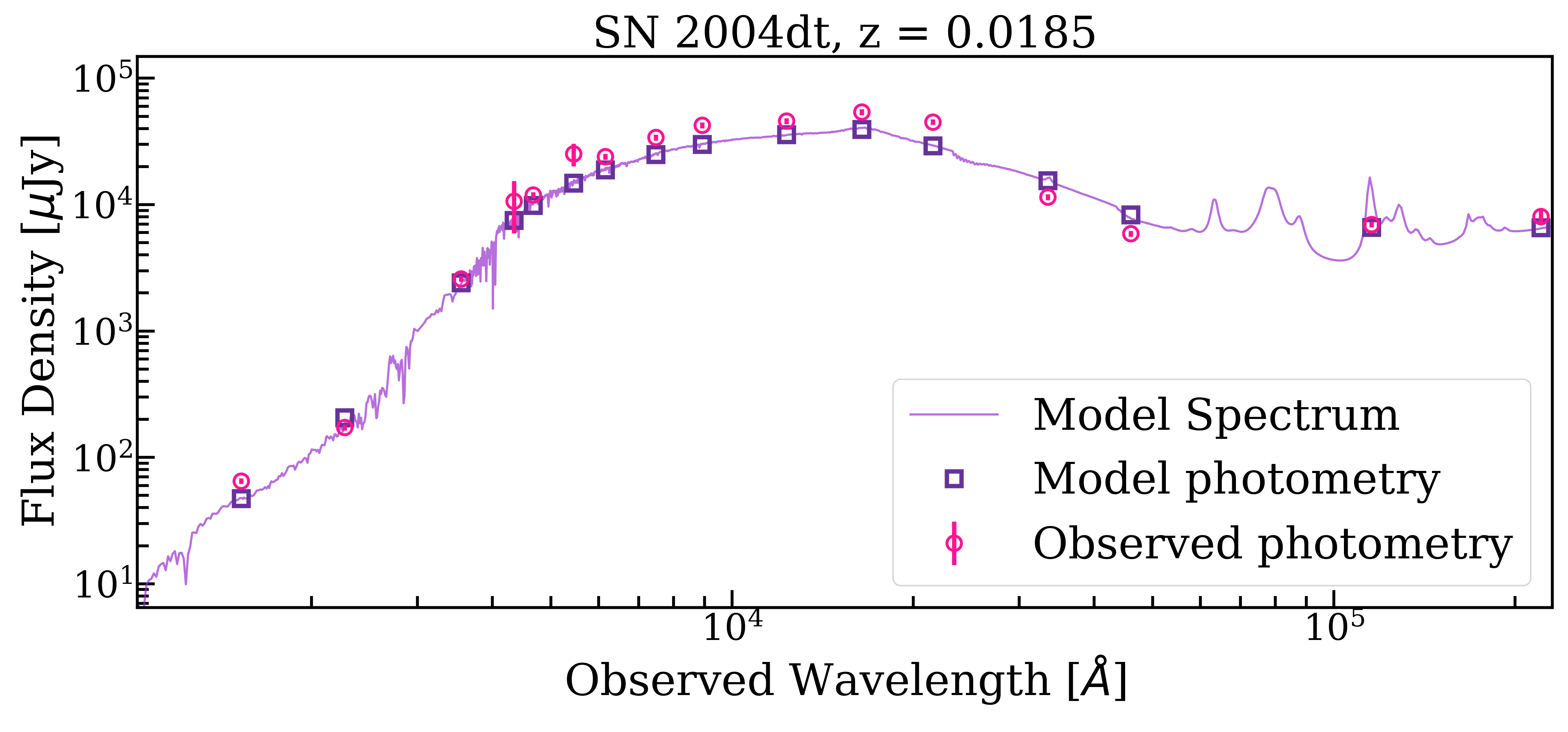}}
\caption{The \texttt{Prospector}-produced model spectra (purple lines) and photometry (purple squares) in comparison to the observed photometry (red circles) for two representative SNe~Ia in our sample: low velocity SN 199dq (left) and high velocity SN 2004dt (right). The model SEDs are produced at the median of each hosts' posterior distributions for their stellar population properties (Table \ref{tab:hostdata}). We showcase these fits to highlight the consistency of the model SED with the observed data, which leads to well-defined posterior distributions for their stellar population properties.}
\label{SED}
\end{figure*}

\subsection{Stellar Population Modeling}
\label{sec:prospector}
To determine the stellar population properties of the SNe host galaxies, we use the stellar population modeling code \texttt{Prospector} \citep{Leja_2017, jlc+2021}. We fit the observational data with \texttt{Prospector} through a nested sampling fitting routine, \texttt{dynesty} \citep{Dynesty}, to return posterior distributions of the stellar population properties of interest. \texttt{Prospector} produces model SEDs with \texttt{FSPS} (\texttt{Flexible Stellar Population Synthesis}; \citealt{FSPS_2009, FSPS_2010}) using single stellar models through \texttt{MIST} \citep{MIST} and the \texttt{MILES} spectral library \citep{MILES}. The main fitted parameters are the age of the galaxy at the time of observation (the maximum allowed value is the age of the universe at the SN's redshift), total mass formed, stellar metallicity ($Z_*$), and $V$-band optical depth.

For all \texttt{Prospector} fits, we use the \citet{Chabrier2003} initial mass function (IMF) and a parametric delayed-$\tau$ star formation history (SFH $\propto te^{-t/\tau}$), defined by the $e$-folding time $\tau$, a sampled parameter. We include the effects nebular emission \citep{bdc+2017} and fix the gas phase metallicity ($Z_{\textrm{gas}}$) to solar since we do not use an observed spectrum in the modeling and therefore cannot measure spectral line strengths. We constrain the total mass formed and stellar metallicities through the \citet{gcb+05} mass-metallicity relation to probe realistic masses ranges for a given stellar metallicity. We measure dust attenuation through the \citet{KriekandConroy13} model, which includes a sampled parameter that determines the offset from the \citet{calzetti2000} attenuation curve. Additionally, we allow the fraction of dust attenuated from young to that from old stellar light to be a sampled parameter to create more flexibility in determining the $V$-band optical depth, which hereafter we report as a $V$-band magnitude ($A_V$). 
As the majority of hosts have available 2MASS and WISE data, we include the \citet{DraineandLi07} IR dust emission model, a three-component dust emission model. To balance dimensionality in the model with the available photometry and ensure that that the model is not overfitting the data, we only sample one of the three components: the polycyclic aromatic hydrocarbon mass fraction ($q_\textrm{pah}$). We follow the methods in \cite{nfd+20} to calculate the stellar mass ($M_*$), present-day star formation rate (SFR), and mass-weighted age ($t_m$) for each host. 

Finally, to classify each host by the degree of star formation, we use the definitions in \citet{Tachella2021}, which measures the combination of the specific SFR (sSFR = SFR/$M_*$; yr$^{-1}$) and the Hubble time $t_\text{H}(z)$ at the SN's redshift: $\mathcal{D}(z) = \text{sSFR} \times t_\text{H}(z)$. If $\mathcal{D}(z) > 1/3$, the host is classified as actively star-forming, if $\mathcal{D}(z) < 1/20$ the host is quiescent (no active star formation), and if $1/20 > \mathcal{D}(z) > 1/3$, the host is transitioning from star forming to quiescent. We describe our main results from stellar population fitting in the following section.
\begin{table*}
\scriptsize
\caption{Stellar Population Properties for SNe Host Samples
\label{tab:prop}}
\begin{tabular}{l|ccccccccccc}
\hline \hline
Sample &
SF &
T &
Q &
$z$ &
$t_m$ [Gyr] &
log(M$_*$/M$_\odot$) &
SFR [M$_\odot$/yr] &
log(sSFR) [yr$^{-1}$] &
log(Z$_*$/Z$_\odot$) &
A$_V$ [mag]  \\ \hline \hline
High Velocity & 57\% & 14\% & 29\% & $0.0158^{+0.0094}_{-0.0095}$ & $2.95^{+6.88}_{-2.49}$ & $10.14^{+0.61}_{-0.5}$ & $0.49^{+5.9}_{-0.48}$ & $-10.17^{+1.08}_{-2.31}$ & $-0.82^{+0.99}_{-0.3}$ & $0.55^{+0.9}_{-0.46}$ \\
Low Velocity &  70\% & 5\% & 25\% & $0.0122^{+0.0115}_{-0.0063}$ &$1.19^{+4.86}_{-0.75}$ & $9.99^{+0.64}_{-0.95}$ & $0.37^{+4.31}_{-0.35}$ & $-10.09^{+1.21}_{-2.35}$ & $-0.83^{+0.8}_{-0.34}$ & $0.55^{+1.58}_{-0.47}$ \\
Faint & 0\% & 25\% & 75\% & $0.0041^{+0.0072}_{-0.0014}$ & $5.74^{+4.78}_{-4.54}$ & $10.1^{+0.65}_{-0.6}$ & $0.01^{+0.03}_{-0.01}$ & $-12.06^{+1.15}_{-3.26}$ & $-0.98^{+1.25}_{-0.19}$ & $0.07^{+0.02}_{-0.05}$ \\ \hline \hline
\end{tabular}
\tablefoot{The star-forming (SF), transitioning (T), and quiescent (Q) fractions and median and the 16$^\textrm{th}$-84$^\textrm{th}$ percentile region for the redshift and \texttt{Prospector}-derived stellar population properties of the hosts of SNe divided into three categories: high velocity, low velocity, and faint SNe.}
\end{table*}

\section{Global Stellar Population Properties}
\label{sec:results}
Here, we discuss the results from the \texttt{Prospector} fitting of 74 SNe hosts. In Table \ref{tab:prop}, we list the median and the 16$^\textrm{th}$-84$^\textrm{th}$ percentile region for the stellar population age, stellar mass, SFR, specific SFR, stellar metallicity, and $A_V$ for the different types of Ia~SNe explored in this paper. We show representative model SEDs in comparison to the selected observed photometry for a high (SN 2004dt) and a low (SN 1999dq) velocity SNe~Ia in Figure \ref{SED} to highlight the accuracy of our \texttt{Prospector} stellar population modeling techniques. We discuss correlations between the \texttt{Prospector}-derived stellar population properties in the following subsections. Individual host properties are shown in Table \ref{tab:hostdata}.

\begin{table*}
\scriptsize
\caption{Anderson Darling Testing Results
\label{tab:ad_tes}}
\begin{tabular}{l|cccccccccc}
\hline \hline
Sample &
$z$ &
$t_m$ &
log(M$_*$/M$_\odot$) &
SFR &
sSFR &
log(Z$_*$/Z$_\odot$) &
A$_V$ &
Offset &
Na I D \\ \hline \hline
High \& Low & 0.25 & 0\% ($0.22$) & 0\% ($>0.25$) & 0\% ($>0.25$) & 0\% ($>0.25$) & 0\% ($>0.25$) & 0\% ($>0.25$) & $0.003$ & 27\% ($0.06$) \\
High \& Faint & 0.02 & 0\% ($>0.25$) & 0\% ($>0.25$) & 76\% ($0.03$) & 80\% ($0.1$) & 0\% ($>0.25$) & 89\% ($0.01$) & $0.22$ & 53\% ($0.003$) \\
Low \& Faint & 0.05 & 16\% ($0.06$) & 0\% ($>0.25$) & 96\% ($0.02$) & 97\% ($0.04$) & 0\% ($>0.25$) & 100\% ($0.0$) & $>0.25$ &  13\% ($0.01$) \\ \hline \hline
\end{tabular}
\tablefoot{The results of Anderson Darling testing of the stellar population properties, offsets, and Na I D equivalent widths for pairs of SNe types in this study (high and low velocity SNe, high velocity and faint SNe, and low velocity and faint SNe). We list the percentage of the 5000 tests that reject the null hypothesis ($P_{AD} < 0.05$) for the stellar population properties and Na I D equivalent widths, and within the parentheses we put the $P_{AD}$ value using distributions constraining the median values for each SN~I.  We list the $P_{AD}$ values for the offsets as there are no uncertainties.}
\end{table*}

\subsection{Redshift}
\label{sec:redshift}
We first focus on the redshift distributions of the three SNe~Ia sub-classes to determine if there are any differences that may bias any of the conclusions in this work, as several stellar population properties (such as age, stellar mass, and SFR) are known to be correlated with redshift. We list the redshift median and the 16$^\textrm{th}$-84$^\textrm{th}$ percentile regions for the three sub-classes in Table \ref{tab:prop}. We compare the distributions through an Anderson Darling (AD test), with the null hypothesis that the hosts are all derived from the same redshift distribution. If the AD test returns a probability $P_{AD} < 0.05$, we can reject the null hypothesis. We list the results of the AD test in Table \ref{tab:ad_tes}. We find no evidence that high and low-velocity SNe~Ia have different redshift distributions. We do, however, find that the the slightly lower median of the faint SNe~Ia does result in a rejection of the null hypothesis between high velocity and faint SNe~Ia, but not between low-velocity and faint SNe~Ia. Given that the majority of our analysis is between high and low velocity SNe~Ia, our AD tests show that there will be no redshift-bias on the stellar population properties for these sub-classes.

\subsection{Star Formation}
\label{sec:starformation}

\begin{figure}[t]
\centering
\includegraphics[width=0.5\textwidth]{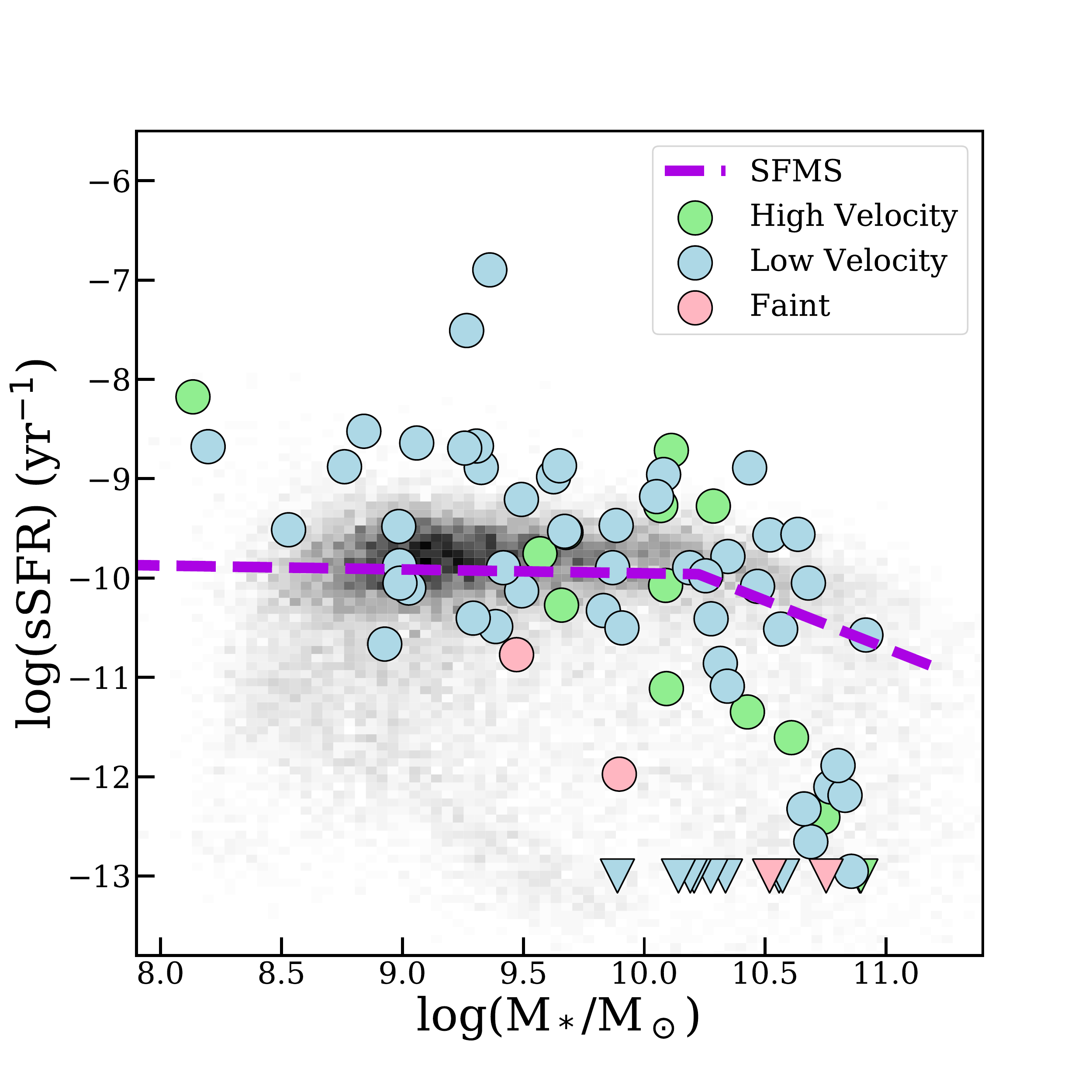}
\vspace{-0.4in}
\caption{The sSFRs and stellar masses for the hosts of SNe divided into three categories: high velocity (green), low velocity (blue), and faint SNe (pink). Downward arrows represent upper limits. We find that both high and low velocity explosions populate nearly the entire SFMS  (purple line; \citealt{Leja2022}), implying their progenitors have breadth of possible host environments with an array of connections to recent star formation activity. We plot the COSMOS2015 field galaxy population \citep{Laigle2016} in the background to show the scatter along the SFMS.}
\label{sfms}
\end{figure}

\begin{figure*}[t]
\includegraphics[width=0.9\textwidth]{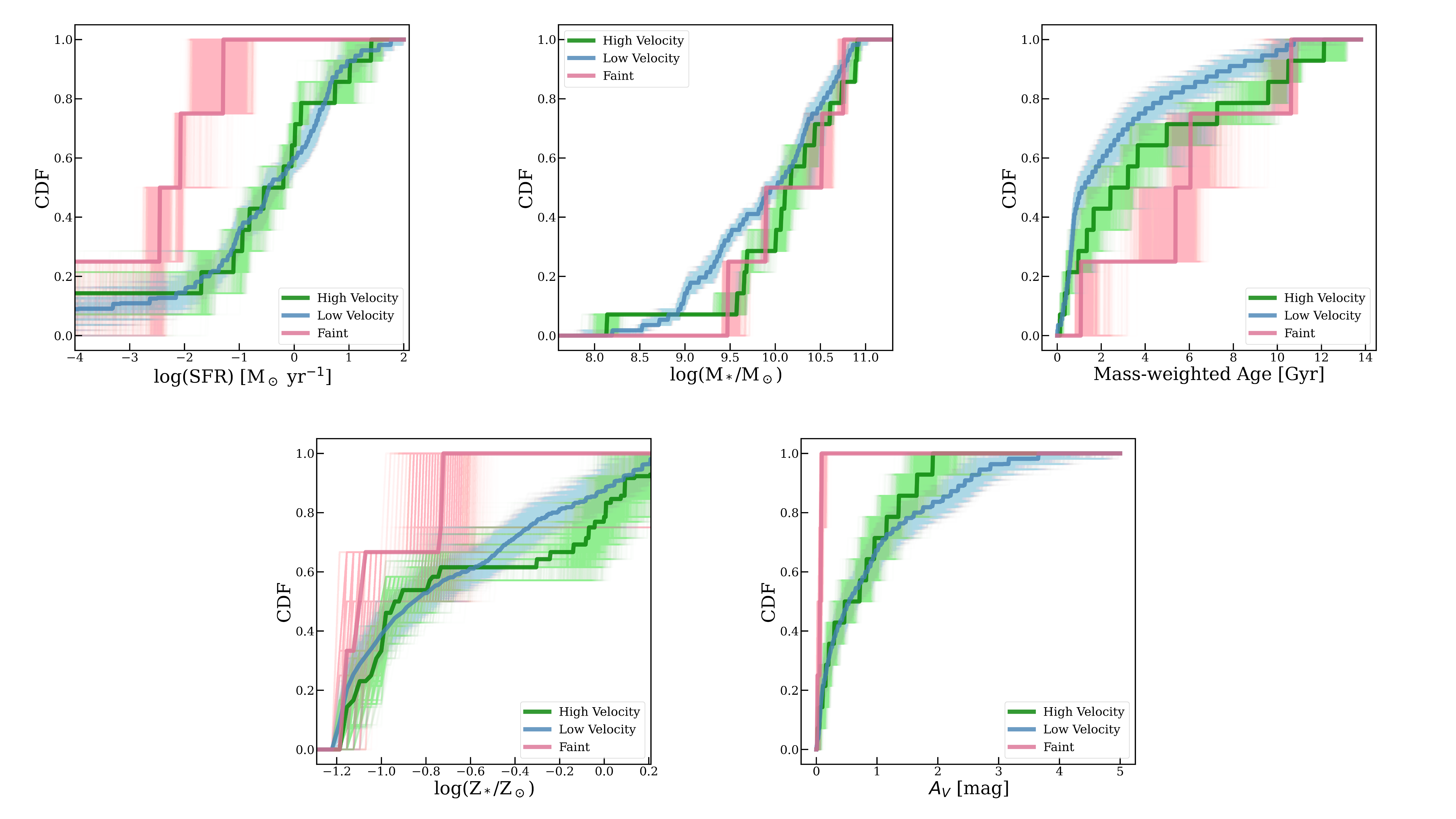}
\caption{\textit{From left to right, top to bottom:} The cumulative distribution (CDF) and 5000 realizations on the CDF for the SFRs, stellar mass (log($M/M_\odot$)), mass-weighted age, stellar metallicity (log($Z_*/Z_\odot$)), and total $A_V$ of high velocity (green), low velocity (blue), and faint SNe (pink) hosts. The darker colors represent the median CDF. We find that there is very little difference in stellar population properties between the high and low velocity SNe~Ia hosts, suggesting that global host properties alone cannot be used to separate these SN classes or distinguish different progenitor types.}
\label{stellar_pop_cdf}
\end{figure*}

We next focus on the global star formation properties of the host galaxies to determine if the SN type is related to star formation activity. We highlight the SNe categories with respect to their peak $M_B$ and \ion{Si}{2} velocities along with their hosts' star formation classification in Figure \ref{sf_class}. We find that $64\%$ of the entire SNe host population are star-forming galaxies, with $28\%$ quiescent galaxies, and $8\%$ transitioning. This roughly follows the star-forming fraction of the observed Type Ia SNe host population \citep{mms+2008}. We list the star-forming, transitioning, and quiescent fraction for the three SNe~Ia sub-classes in Table \ref{tab:prop}. The high and low velocity SNe~Ia hosts roughly follow the fractions of the entire population and the faint SNe hosts are dominated by quiescent galaxies. To test whether the host populations of high and low velocity SNe~Ia are statistically different, we perform a $\chi^2$ contingency test using the Python function in the \texttt{scipy} package. We assume a null hypothesis that high and low velocity SNe~Ia have consistent star forming, transitioning, and quiescent fractions. We determine a $p$-value of 0.45, which is much greater than the result needed to reject the null hypothesis ($<0.05$). We further perform the same tests for both populations in comparison to the full SNe host sample, finding p-values $> 0.70$ for both tests. Thus, we determine that the respective star-forming fractions of the high and low velocity SNe~Ia hosts are consistent with each other and with the general population. 

We further analyze the hosts by comparing their specific star formation rates (sSFRs) versus stellar mass. We compare how these properties populate the star-forming main sequence (SFMS), a well-studied galaxy relation that tracks the SFRs of star-forming galaxies as they gain stellar mass \citep{whitaker2014, speagle2014, Leja2022}. How transient hosts track the SFMS has important implications for the environmental conditions their progenitor traces (e.g. dependencies on bursts of star formation or the amount of stellar mass). In Figure \ref{sfms}, we plot the sSFRs and stellar masses of the SNe hosts in each of the three categories, along with the SFMS derived in \citet{Leja2022} for $z<0.5$. This SFMS was derived from a population of galaxies in the COSMOS-2015 and 3D-HST surveys with stellar population modeling fitting from \texttt{Prospector}, thus is comparable to our results \citep{Leja2022}.  We choose to use the sSFR as opposed to the SFR, as it normalizes the amount of star formation per stellar mass unit, which is useful when comparing galaxies across large stellar mass ranges. We find that both the high and low velocity SNe~Ia populate the entire SFMS and are not clustered into any particular combination of sSFR and stellar mass. This, thus, implies that both of their progenitors depend on a wide array of environmental factors, suggesting a breadth of possible host environments and connections recent star formation activity. Interestingly, for both host samples, we do find that at $\lesssim 10^9 M_\odot$, all hosts lie above the SFMS. It is not clear whether this is an artifact of the galaxy-targeted nature of the LOSS survey or a real progenitor effect (i.e., high and low velocity SNe~Ia progenitors are not as capable of forming in low mass, low sSFR galaxies). While it is a small sample size, our faint SNe sample all fall below the SFMS and reside in higher stellar mas galaxies. If representative of the entire faint SNe sample, this would imply that the faint SNe progenitor is more apparently connected to stellar mass rather than star formation (and therefore likely has an older stellar progenitor).

We next compare cumulative distribution functions (CDFs) of SFR between the three categories of SNe and their hosts. We build the CDFs by randomly drawing 5000 values from each host's \texttt{Prospector}-derived posterior distribution for each property. We note that there is no change in the results if we increase beyond 5000 draws. We then build 5000 CDFs for each SN category, as shown in Figure \ref{stellar_pop_cdf}. We compare all CDFs through an AD test, with the null hypothesis that the hosts are all derived from the same SFR distribution, and calculate the percentage of tests with a probability $P_{AD} < 0.05$ and list these values in Table \ref{tab:ad_tes}. We also list the $P_{AD}$ determined from distributions built from the median SFR of each SN in Table \ref{tab:ad_tes}. We choose these comparison methods to encapsulate the uncertainty in the posterior distributions for each SN~Ia sub-class. However, we caution that this will not take any uncertainty from the sample sizes of each SN~Ia sub-class into account, which may be important in the case of the faint SNe~Ia that have a significantly smaller sample size than the other two sub-classes. We find that when comparing the SFRs of high and low velocity SNe~Ia hosts, $0\%$ of AD tests result in a rejection of the null hypothesis, with the same result being found when comparing sSFRs. 
We thus infer that if the high and low velocity SNe~Ia~Ia represent different progenitors, neither progenitor is more dependent on the amount of recent star formation in their environment and thus, the amount of global star formation in a host cannot be used to distinguish these SN populations. Meanwhile, the null hypothesis can be rejected in $76\%$ ($80\%$) and $96\%$ ($97\%$) of the SFR (sSFR) tests for the faint SNe and high velocity SNe~Ia hosts (faint SNe and low velocity SNe~Ia hosts).
This implies that that the faint SNe progenitor is less dependent on recent star formation than either the high or low velocity SNe~Ia, if the population studied here is representative of the full faint SNe population.  Overall, these results point to the fact that we cannot separate high and low velocity SNe~Ia from the amount of global star formation in their hosts, although we note that locally they may indeed trace different star formations.

\subsection{Stellar Mass, Age, Metallicity, and Dust}
\label{sec:stellarpop}
\begin{figure*}[t]
\centering
\includegraphics[width=1.0\textwidth]{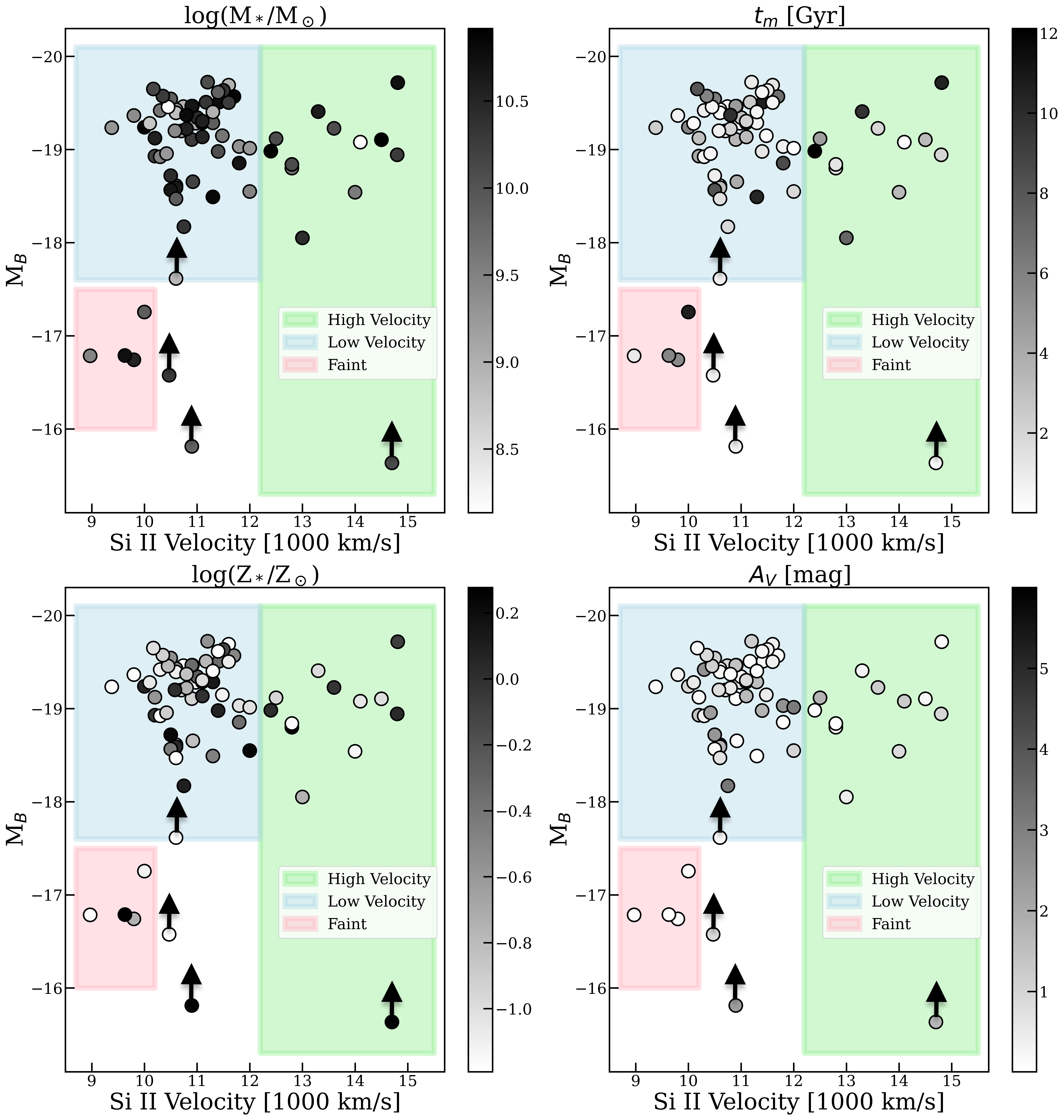}
\caption{The same plot as in Figure \ref{sf_class}, with SNe colored by their hosts' amount of stellar mass (log($M/M_\odot$), top left), mass-weighted age ($t_m$, top right), stellar metallicity (log($Z_*/Z_\odot$), bottom left), and total $A_V$ (bottom right). 
We find that there appears very little correlation between host stellar population property and SN property.}
\label{stellar_pop}
\end{figure*}

We next compare the main stellar population properties (stellar mass, stellar metallicity, stellar age, total dust) of each SN group's hosts. We show the $M_B$ vs. \ion{Si}{2} velocity relation of the SNe colored by the median stellar mass, stellar population age, stellar metallicity, and dust from the \texttt{Prospector} posterior distributions of their hosts in Figure \ref{stellar_pop}. We list the derived median and the 16$^\textrm{th}$-84$^\textrm{th}$ percentile region for each SN category in Table \ref{tab:prop}. We find that there is very little difference in stellar population property compared to SN type. However, as the faint SNe are almost exclusively in non-star forming hosts, they reside in older galaxies with lower dust and higher stellar mass than the hosts of the other SN types. Furthermore, we see little evidence that the peak $B$-band luminosity and \ion{Si}{2} velocities are correlated at all with host galaxy stellar population properties.

In Figure \ref{stellar_pop_cdf}, we compare the CDFs of each stellar population property for the three SN categories (see Section~\ref{sec:sample}) and compare the distributions through AD tests, results of which are listed in Table \ref{tab:ad_tes}. We cannot reject the null hypothesis for any of our comparisons in stellar mass and stellar metallicity. This implies that all SNe samples are consistent with having common stellar mass and stellar metallicity distributions. This consistency in the stellar mass distributions between high and low velocity SNe~Ia hosts deviates from the conclusion found in \citet{pan2020}, in which all high velocity SNe~Ia (defined as having Si II velocities $> 12,000$~km~s$^{-1}$, at $z < 0.19$ were found in galaxies with $>10^{9.6}M_\odot$ and preferentially occurred in more massive galaxies than low velocity SNe~Ia. Similar results to \citet{pan2020} were found in \citet{djd+2021}. In this work, we find one high velocity SN~Ia host lower than this stellar mass minimum, at $\approx10^{8.1}M_\odot$ (the host of SN 2007qe, which was not included in the \citealt{pan2020} sample and had no stellar mass estimate in \citealt{djd+2021}), thus this is likely driving the stellar mass distributions to be more similar. We also note that the stellar mass estimates determined in this work are all $\approx 0.3$~dex smaller than the masses determined in \citet{pan2020} between the 10 shared hosts, which is a known effect when switching from a Salpeter IMF (used in \citealt{pan2020}) to a Chabrier IMF (this work). However, most importantly, the difference will only affect individual stellar mass measurements and not how the distributions of stellar masses compare. Thus, the similarity of the stellar mass distributions between high and low velocity SNe~Ia found in this work is real, and not simply caused by a different stellar population modeling technique or SED fitting tool. Interestingly, the \citet{pan2020} and \citet{djd+2021} studies are both based on larger SNe host samples, from e.g., the Foundation Supernova Survey \citep{fcr+2018} and the PTF survey, which should better target SNe in lower mass galaxies than the LOSS survey used in this work \citep{psm+2015, graur2017}. Thus, this current discrepancy highlights the need for further, uniform host studies with larger SN samples to more robustly determine whether high velocity SNe~Ia do occur significantly in lower mass galaxies. At this time, our results may simply show that high velocity SNe~Ia can occur in a more diverse array of environments than previously known.

For mass-weighted age, we can reject $0\%$  of tests between high and low velocity SNe~Ia hosts and $16\%$ of tests between low velocity and faint SNe hosts. The null hypothesis cannot be rejected in any of the trials between high velocity and faint SNe hosts. Furthermore, we find that $0\%$ of tests can be rejected between $A_V$ distributions between the high and low velocity SNe~Ia hosts. However, as the faint SNe hosts tend to have very little dust, we find that $89\%$ of AD tests can be rejected between faint SNe and high velocity SNe~Ia host $A_V$ distributions, and $100\%$ of tests can be rejected between faint SNe and low velocity SNe~Ia host $A_V$ distributions. We do note, once again, that the faint SNe sample size is small, thus these conclusions are only substantiated if they are representative of their entire population.

Overall, we find very little evidence for significant stellar population property difference between high and low velocity SNe~Ia populations, suggesting the global environment alone cannot distinguish the two SNe and that their progenitors' environments have similar stellar population properties. Furthermore, this suggests that the differences in SN properties of either group cannot be explained by differences in their global host environments. We find more substantial evidence that the environments of faint SNe are quite different, however note that the population of faint SNe we have explored in this paper is quite small thus might not be fully representative of the host properties of all faint SNe.

\section{Evidence for Difference in Local Environments}
\label{sec:local}
As discussed in the previous section, we find little evidence that the global environmental properties of possible high and low velocity SNe are different. In this section, we discuss if their progenitors are tracing different local environments through analyzing their observed physical offset distributions and Na I~D equivalent widths.

\subsection{Offsets}
\label{sec:disc_offsets}

\begin{figure}[t]
\centering
\includegraphics[width=0.45\textwidth]{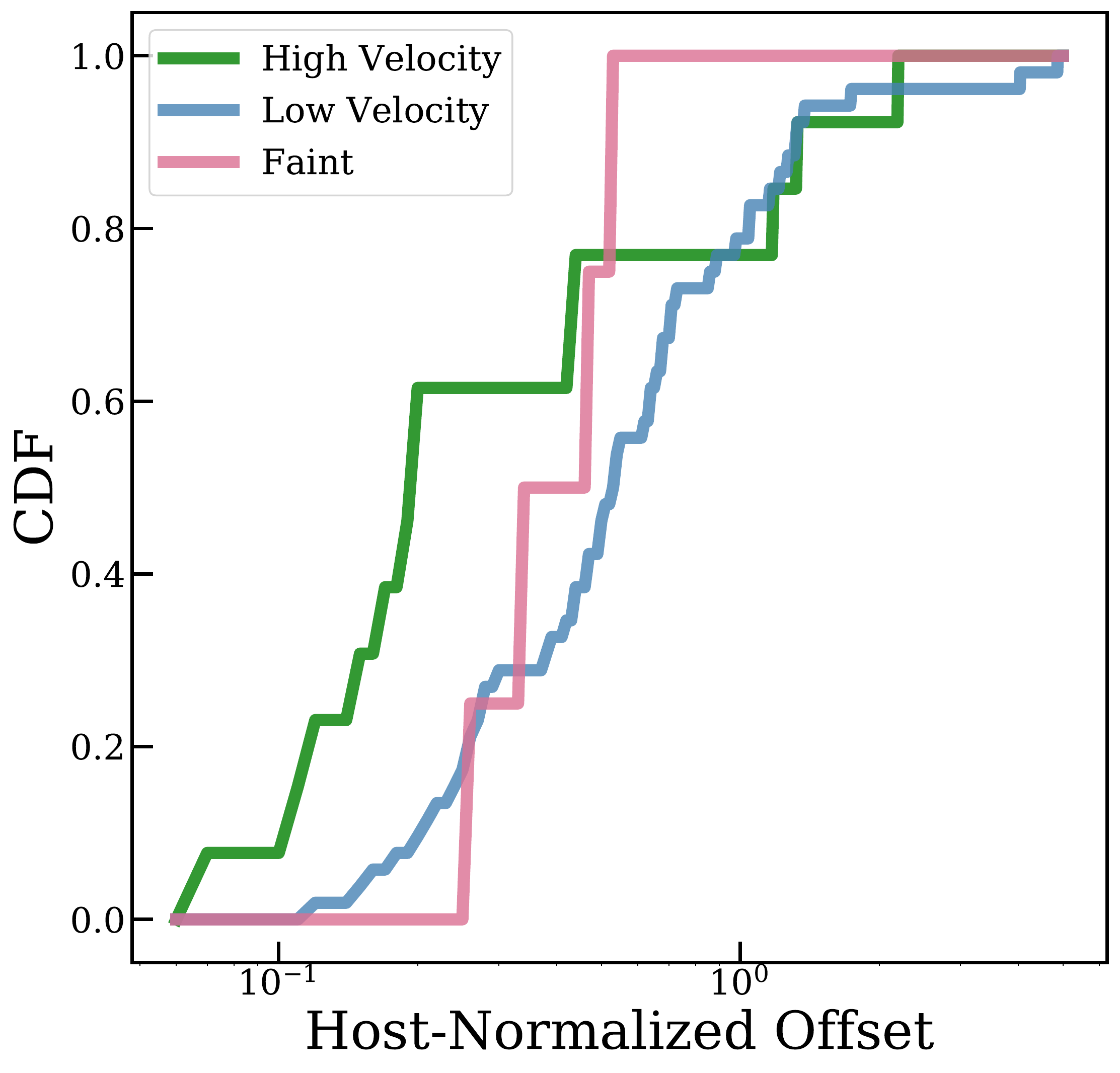}
\caption{CDFs of the host-normalized offsets (physical offset of SNe from the center of the host normalized by the $K$-band total radius) for the high velocity (green), low velocity (blue), and faint SNe (pink) classifications. High velocity SNe~Ia have smaller host-normalized offsets than either low velocity of faint SNe, implying they lie more towards the center of their hosts than the other SN sub-classes. This may be indicative of a distinct local environment.}
\label{offsets}
\end{figure} 

The offsets of transients from the center of their host galaxies has been used to probe local environments, migrations from progenitor birthplaces, and other key factors that help distinguish transient types and progenitors (e.g. \citealt{Kasliwal2012, Blanchard2016, Fong2022}). Thus, we determine if the SNe~Ia sub-classes studied here have distinct offset distributions. We obtain angular separations of our SNe sample to the center of their hosts and the 2MASS K-band ``total" radii of the host galaxies, which is available for 69 out of the 74 host studied in this work (13/14 high velocity SNe~Ia and 52/56 of the low velocity SNe~Ia hosts) from NED. The 2MASS K-band ``total" radius is defined as the isophotal radius (the semi-major axis at 20 mag/arcsec$^2$ isophote at Ks) added to the integration of the surface brightness profile that extends from the isophotal aperture out to $\sim4$ disk scale lengths. This radius is typically 10-20\% larger than the isophotal radius \citep{2MASS}. We determine host-normalized offsets, or the relative location of the SN within its host, by dividing the angular separation of each SNe to their hosts' centers by the host galaxies' size. We find a median and the 16$^\textrm{th}$-84$^\textrm{th}$ percentile region for host-normalized offsets to be $0.19^{+0.99}_{-0.08}$~$r/r_e$ for high velocity SNe~Ia hosts,  $0.53^{+0.61}_{-0.29}$~$r/r_e$ for low velocity SNe~Ia hosts, and $0.4^{+0.1}_{-0.11}$~$r/r_e$ for faint SNe, and we show the respective CDFs in Figure \ref{offsets}. When conducting an AD test between pairs of the three SNe offset distributions, we find that we cannot reject the null hypothesis for any pair ($P_{AD} \geq 0.20$) except the high and low velocity SNe~Ia ($P_{AD} = 0.003$), as high velocity SNe~Ia are more concentrated towards the center of their hosts than low velocity SNe~Ia. These results are in agreement with those in \citet{wang2013} (52\% of our SNe sample overlaps with this sample), which studied the relative locations of high and low velocity SNe~Ia within their hosts, finding high velocity SNe~Ia occur at lower radial distances. Future studies focused on spectral emission at these locations (such as star formation tracers like $H\alpha$) may reveal if these offsets are indicative of a local stellar population property difference. Furthermore, the difference in offset distributions could indicate that high velocity SNe~Ia appear redder from extinction in their local environments, as more gas and dust lies towards the center of the galaxies. Overall, these results highlight that while global properties of high and low velocity SNe~Ia may not be distinguished, the local environments may play a larger role in separating these sub-classes.

\begin{figure}[t]
\centering
\includegraphics[width=0.49\textwidth]{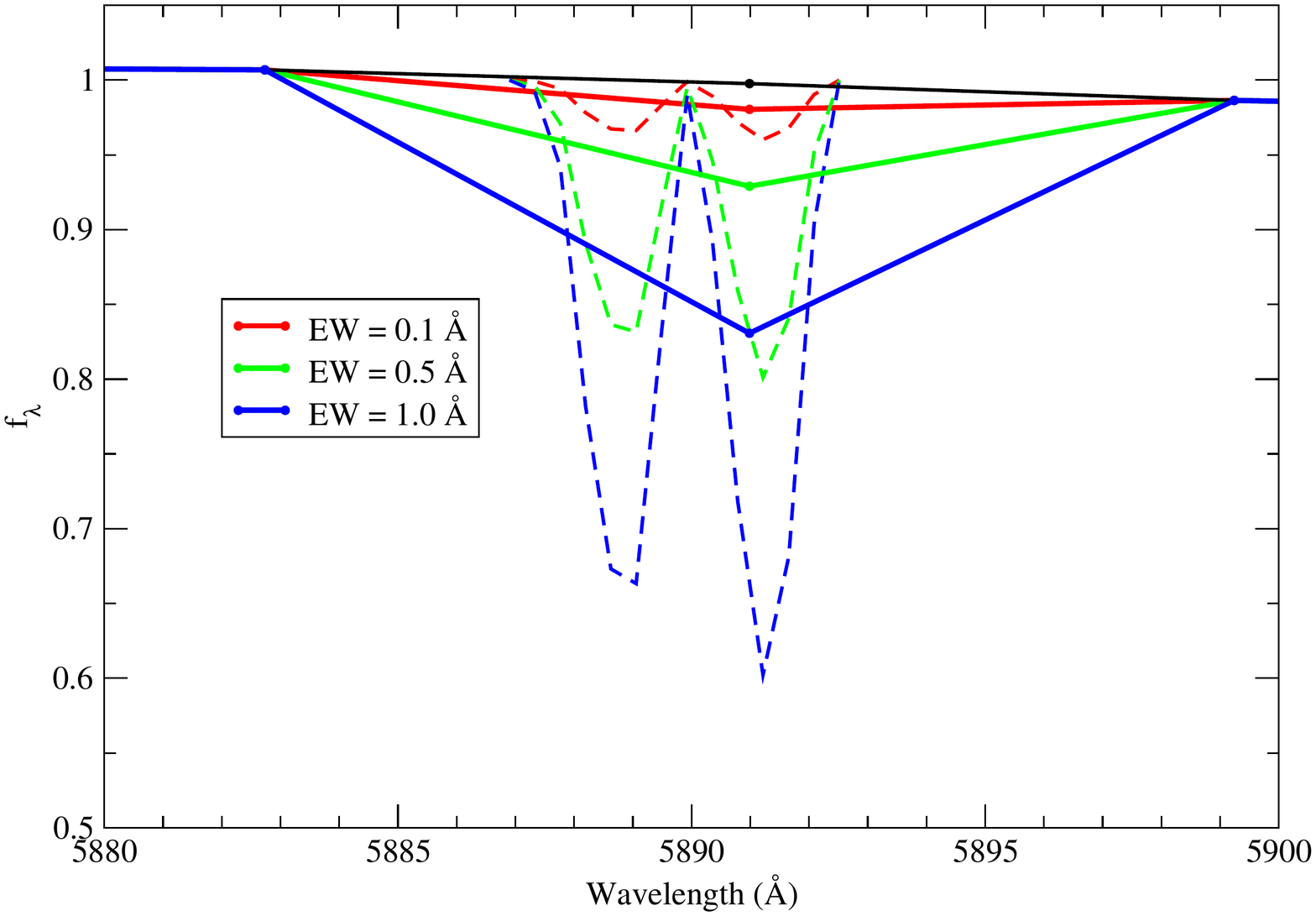}
\vspace{-0.3in}
\caption{An example measurement of the equivalent width on the day~$+$4 spectrum of SN~1990N  (one of the lowest resolution spectra in our sample; \citealt{1991ApJ...371L..23L}) seen in black. An upper limit of 0.1~\AA\ can be placed on this individual spectrum. We present example strengths of the lines at 0.1, 0.5 and 1.0~\AA\ in both high-resolution (dashed) and at the resolution these Na~I~D lines would have in the lower resolution of the observed spectrum of SN 1990N (in solid  red, green, and blue respectively). The noise for each measurement is taken directly from that reported by the observer or estimated from the signal-to-noise of the continuum around this feature. In the above case, the error bars are equaly to the line width, $S/N \sim 100$ per resolution element, which is the typical for many of these nearby SNe~Ia. All measurements include the uncertainty in the continuum which is very small given the short wavelength range covered by these features.}
\label{NaID_measure}
\end{figure}

\begin{figure}[t]
\centering
\includegraphics[width=0.49\textwidth]{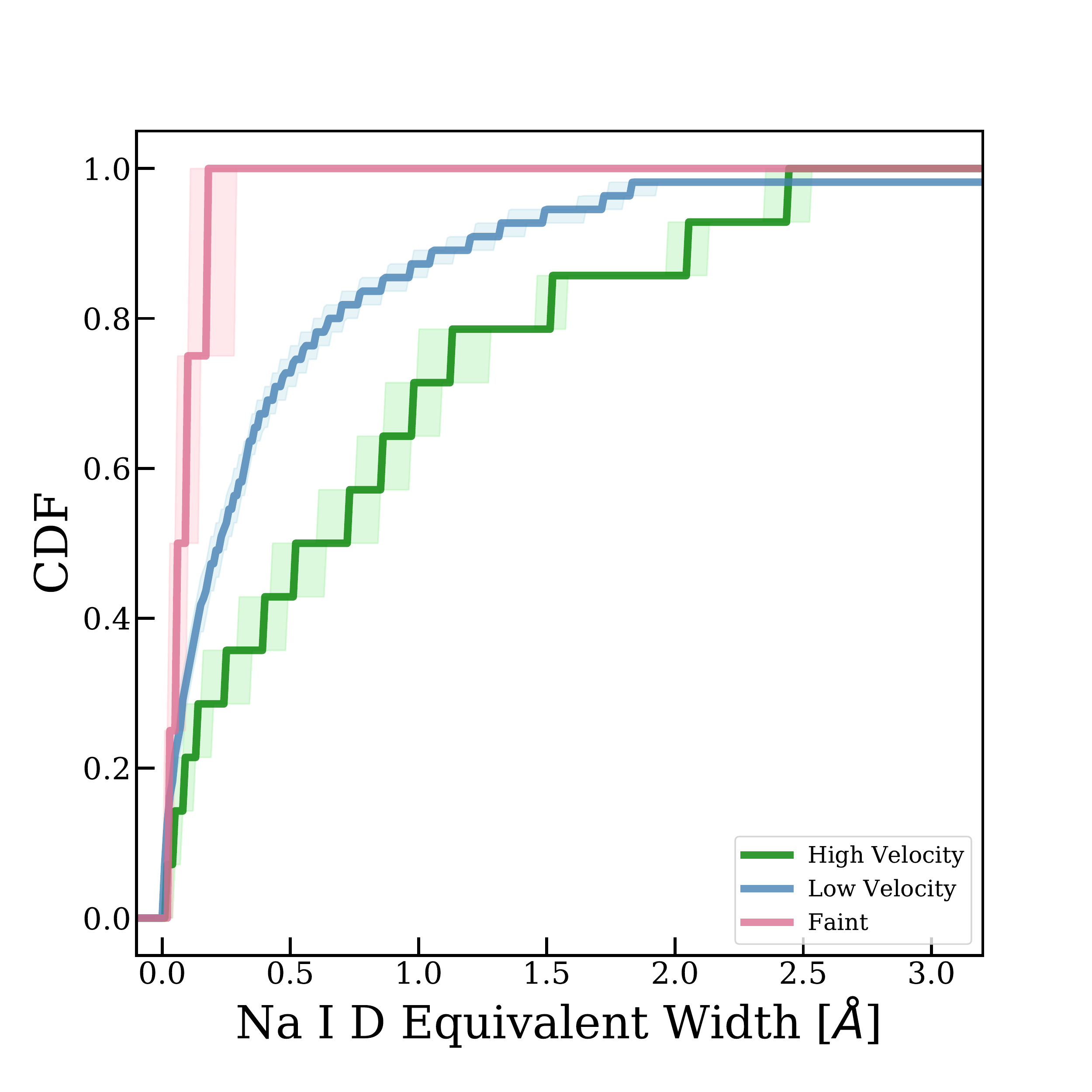}
\vspace{-0.3in}
\caption{CDFs on the Na I~D line equivalent widths for high velocity (green), low velocity (blue), and faint (pink) SNe. Darker lines represent the median of the distribution and lighter lines represent uncertainty on the CDF. We cannot reject the null hypothesis that high and low velocity SNe~Ia have similar Na~I~D equivalent width distributions with AD statistical tests.}
\label{NaID}
\end{figure}

\subsection{Na I~D Equivalent Widths}
\label{sec:NaID}
Finally, we analyze the equivalent widths of the Na I~D lines in each SN spectrum to infer if high and low velocity SNe~Ia probe different circumstellar or interstellar environments. The Na I~D line strength probes the host dust and circumstellar gas, dust, and metals \citep{Blondin2009, folatelli2010, poznanski2012, phillips2013}, thus is a good indicator of the local environment surrounding the SN \citep{sgs+2011,sgs+2014}. We collect published Na I~D equivalent widths for 61 SNe in \citet{wang2019} and references therein. For the 13 remaining SNe: 1990N \citep{jlk+1992, gl1990}, 1991bg \citep{frb+1992}, 2005hk \citep{plf+2007}, 2000E, 2006ax, 2006lf, 2008Q \citep{bmk+2012}, 2000dr, 2002cs, 2003Y, 2003gn, 2003gt, 2007on \citep{sff+2012}, we collect spectra from WISeREP \citep{WISeREP}. To determine the Na~I~D equivalent widths for these SNe~Ia, we performed a standard equivalent width measurement on these mostly lower resolution spectra as seen in Figure~\ref{NaID_measure}. In the case where a SN has multiple high signal-to-noise spectra, we measured the equivalent width on each spectrum and combined the individual equivalent measurements, weighted by their uncertainties, to provide a single equivalent width and uncertainty for each SN. We note that in all but two cases (SNe~2006lf and 2007on), they are upper limits. We determine median and the 16$^\textrm{th}$-84$^\textrm{th}$ percentile regions of the Na I~D equivalent widths for each SN~Ia sub-class. We find that high velocity SNe~Ia have Na I~D equivalent widths $0.62^{+0.95}_{-0.55}$~\AA, low velocity SNe~Ia have equivalent widths $0.22^{+0.61}_{-0.19}$~\AA, and faint SNe have equivalent widths $0.07^{+0.09}_{-0.05}$~\AA. 

\begin{figure}[t]
\centering
\includegraphics[width=0.49\textwidth]{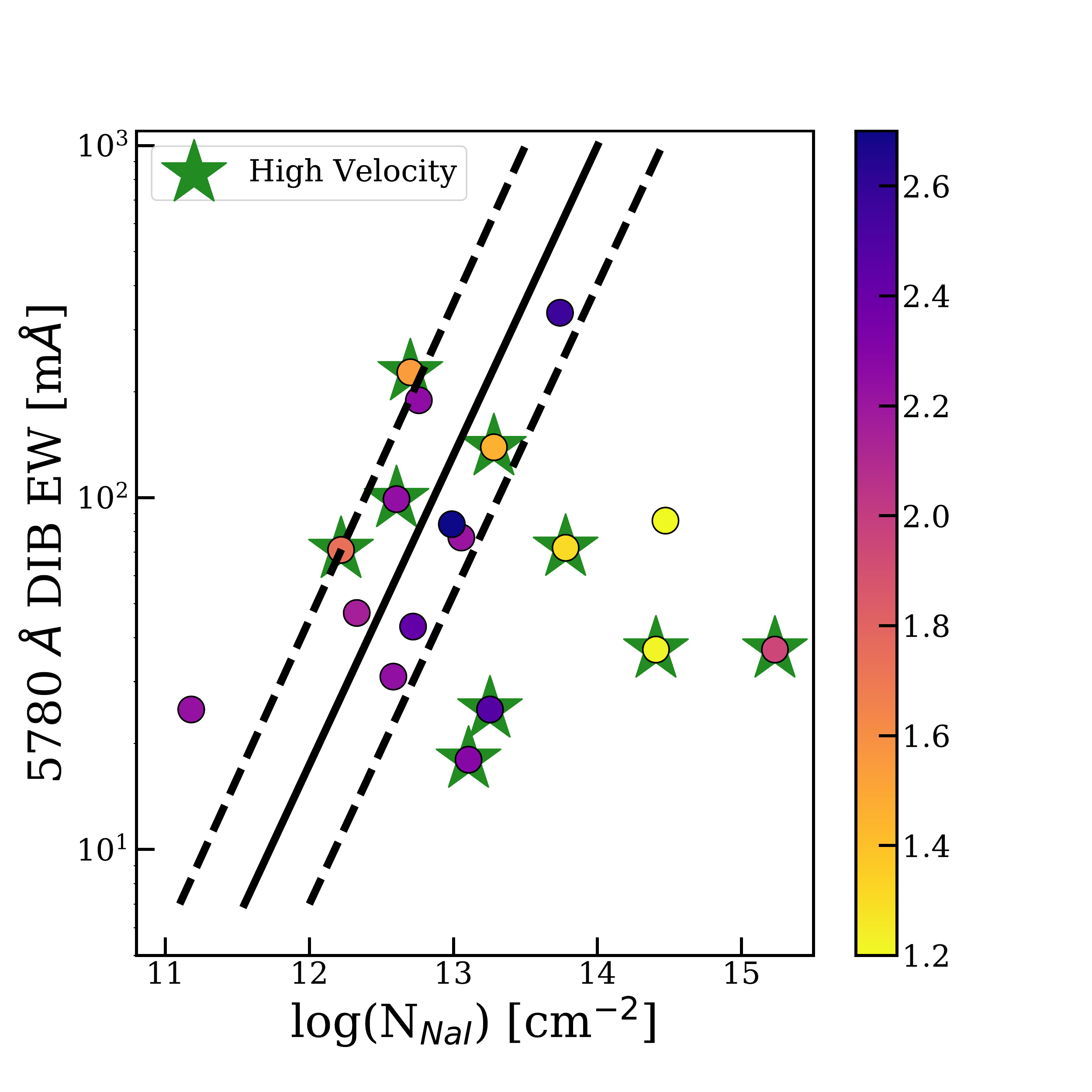}
\caption{The equivalent width of the 5780~\AA\ DIB vs. column depth of Na~I for those SNe~Ia from \citet{phillips2013} that we can separate between low and high velocity. We have color coded the data points by their assumed $R_V$. We note that uncertainties are smaller than symbols. The black line represents the correlation found in the Milky Way galaxy and the dashed black represent the $1\sigma$ uncertainty on the relation. While all but two of the low velocity SNe~Ia fall within the $1\sigma$ region of the Milky Way relation, over half the high velocity SNe~Ia fall outside this correlation.}
\label{NaID_DIB}
\end{figure}

To make CDFs of the Na I~D equivalent widths, we build Gaussian distributions using their median and $1\sigma$ uncertainties, with a minimum possible value of 0~\AA, and sample from this distribution 5000 times. We show the median and the 16$^\textrm{th}$-84$^\textrm{th}$ percentile region for these CDFs in Figure \ref{NaID}. We compare pairs of the 5000 CDFs and list the percentage of tests that reject the null hypothesis in Table \ref{tab:ad_tes}. We find that we can reject the null hypothesis in 27\% of the tests between high and low velocity SNe~Ia, 53\% of tests between high velocity and faint SNe, and 13\% of tests between low velocity and faint SNe. Thus, there is little statistical evidence that high and low velocity SNe~Ia have different Na~I~D equivalent widths.

\section{Discussion}
\label{sec:discussion}
In this work, we have focused on the global host galaxy and local environmental properties of high and low velocity SNe~Ia to see if we can find evidence for an environmental condition that could explain why high velocity SNe~Ia appear redder. While we find no evidence that the global host properties can account for these differences in SN observables, we have shown that more locally, using the host normalized offsets, high and low velocity SNe~Ia environments may indeed differ. We also note that while we did not find statistical support for different Na I~D equivalent width distributions between high and low velocity SNe~Ia, with even a minimally larger high velocity SN sample, these distributions may indeed become distinct. For example, if there were only 1-2 more high velocity SNe~Ia that showed Na I~D equivalent width strengths at the median of the population studied here, $>50 \%$ of Anderson Darling tests would be rejected between high and low velocity SNe~Ia. At the minimum, this signifies a need for larger samples of these SNe~Ia sub-classes from volume-limited surveys that are not biased against finding more extinct, centrally located SNe~Ia, for more robust statistical tests. At most, the potential difference in both host-normalized offsets and Na I~D equivalents could have major implications for whether the environment or the progenitor causes high velocity SNe~Ia to have redder colors, which we briefly discuss below.

Understanding the contribution of the local environmental properties versus the progenitor properties to the redness of high velocity SNe~Ia requires discerning whether the excess Na I~D gas possibly observed in high velocity SNe comes from the circumstellar or interstellar medium. Indeed, \citet{phillips2013}, which explored the properties of SNe~Ia as a function of both the diffuse interstellar bands (DIBs) and Na~I~D lines through high-resolution spectroscopy, found that (i) the low $R_V$ values derived for very reddened SNe~Ia are likely caused by dust in the interstellar medium, and not the circumstellar medium, and (ii) that nearly a quarter of SNe~Ia have large host Na~I column densities in comparison with the amount of dust observed to redden their spectra. \citet{phillips2013} further noted that all of the SNe~Ia with unusually strong Na~I~D lines have ``blueshifted” profiles in the classification scheme following \citep{2011Sci...333..856S} and that the majority of the high velocity SNe~Ia (seven out of ten) have this ``blueshifted" feature. This paper, however, relies on one major assumption: that the intrinsic colors of both the high and low velocity SNe~Ia are similar as defined by the SNooPy algorithm \citep{2011AJ....141...19B}. Here, we revisit these conclusions given more recent work on the potential origins of these two sub-groups of SNe~Ia.

Indeed, there already exists theoretical evidence that high and low velocity SNe~Ia may have different intrinsic colors caused by separate progenitors. For example, \citet{pnk2019} suggests that the redder colors of high velocity SNe~Ia are a natural consequence of a sub-$M_\textrm{ch}$ mass explosion, while the bluer colors of low velocity SNe~Ia are due to classical $M_\textrm{ch}$ explosions. We note that while others claim that the bulk of the low velocity SNe~Ia can also derive from sub-$M_\textrm{ch}$ explosions \citep{shen2021}, we use the \cite{pnk2019} model to guide our discussion and simply note that this relationship implies that we are seeing at least two separate progenitor systems and/or explosion mechanisms with {\it intrinsically} different colors. Nonetheless, this only strengthens the findings in \citet{phillips2013} as the anomalously low values of $R_V$ they found for reddened SNe~Ia would become even lower and the lack of correlation between the strength of the Na~I~D lines and the DIBs, or assumed dust from the underlying colors, would become even smaller. An example of this can be seen in Figure~\ref{NaID_DIB}. Here, we plot the equivalent width of the 5780~\AA\ DIB versus column depth of Na~I for those SNe~Ia from \citet{phillips2013} that we can separate between low and high velocity. While all but one of the low velocity SNe~Ia tightly follow the correlation found in the Milky Way galaxy, nearly half the high velocity SNe~Ia lie outside this correlation. Such a result strongly implies that the dust is a property of the ISM and furthermore that the CSM of these events is almost certainly contributing a clear excess to the Na~I~D profiles \citep{2013MNRAS.436..222M, phillips2013}. \citet{2013MNRAS.436..222M} further found that SNe~Ia with blueshifted Na~I~D profiles had both an excess absorption compared to those with non-blueshifted profiles and redder $B-V$ colors at peak brightness.

While we have only considered three sub-groups of SNe~Ia  with simple velocity and peak brightness cuts, we note that the sub-$M_\textrm{ch}$ models of \citet{pnk2019} extend down to more normal velocities (v $\sim 10,750$~km/s)  and slightly fainter peak absolute magnitudes (B $\sim -18.5$). It would be interesting to see in a future, larger, volume limited, sample if the blueshifted SNe~Ia noted by \citet{2013MNRAS.436..222M} and \citet{phillips2013} shared the characteristics of this sub-population of SNe~Ia. Additional future work is also needed to understand how the more local environments of these sub-$M_\textrm{Ch}$ explosion candidates may further affect their colors and SN observables, as we have noted how high velocity SNe~Ia trend towards smaller offsets from their hosts' centers and are thus likely more observationally affected by extinction due to their more central locations.

\section{Conclusions}
\label{sec:conclusion}
In this study, we have determined the host galaxy stellar population properties for 74 SNe~Ia at $z < 0.04$, divided into the sub-categories of high velocity, low velocity, and faint SNe~Ia based on their peak $B$-band luminosities and \ion{Si}{2} velocities. All SNe~Ia have good photometric coverage for determination of their properties and their host galaxies have at least three bands of archival photometry, the minimum requirement for a stellar population fit. We summarize our main conclusions below:
\begin{itemize}
    \item We find that there is no statistical evidence for differences between the high velocity (\ion{Si}{2} velocities $\gtrsim 12,200$~km~s$^{-1}$) and low velocity SNe~Ia host galaxy populations in SFR, sSFR, stellar mass, stellar population age, dust attenuation, and stellar metallicity. Within the small sample used in this paper, the fainter ($M_B > -17.5$~mag) SNe occur in older, more massive, and more quiescent galaxies than the rest SNe population.
    \item In our comparison between the host-normalized offsets for the SNe (from SN position to center of the host), we find high velocity SNe~Ia are more concentrated towards the center of their hosts than low velocity SNe~Ia. This suggests that there may be local environmental differences between these populations that are not distinct in their global properties.
    \item Finally, we find that high velocity and low velocity SNe~Ia do not have statistically different Na~I~D EW distributions, implying there is little evidence that surrounding Na gas is effecting the observable differences of these populations.
\end{itemize}

Our results on the stellar population properties of SNe~Ia hosts strongly suggest that the global environments of high and low velocity SNe~Ia are not intrinsically different. While this might be a cause of the galaxy-targeted nature of the SNe~Ia survey we used, which generally has more limitations in the host types (e.g, more massive galaxies; \citealt{psm+2015}), we have also shown that high velocity SNe~Ia are capable of occurring in more diverse and less massive hosts than studies focused on SNe~Ia samples without this survey bias \citep{pan2020, djd+2021}. We do, however, find that the smaller host-normalized offsets of high velocity SNe~Ia may show that the local environments between these populations are different. Future work necessitates the use of high-resolution photomotery and/or Integral Field Unit (IFU) spectroscopy to resolve the local environments of nearby high velocity SNe~Ia to more conclusively determine if these offsets are indicative of separate environmental properties than more normal velocity SNe~Ia and more conclusively inform if the  Na I D gas in high velocity SNe~Ia is interstellar or circumstellar. Indeed, \citet{rigault2020} has already shown that the local star formation activity surrounding SNe~Ia does affect the corrections to their peak magnitudes used in the standardization process for cosmology. Furthermore, it has been proposed that sub-$M_\textrm{ch}$ progenitors are likely more connected to star formation and occur in younger stellar population than classical SNe~Ia \citep{dkss2020}. Thus, if high velocity SNe~Ia do trace younger and more star-forming regions within their hosts, it may be indicative of their possible sub-$M_\textrm{ch}$ progenitor, as proposed by \citet{pnk2019}.

Our conclusions notably highlight that the global stellar population properties cannot be used to separate classes of SNe~Ia for future cosmology measurements. This is especially problematic for the upcoming era of the Vera Rubin Observatory (VRO), in which hundreds of thousands of SNe~Ia will be discovered, the majority of which will not be followed up with spectroscopy of the supernova to determine their \ion{Si}{2} velocities. Furthermore, if the local environments of high velocity SNe are indeed different then these environmental differences may only be probed for more local universe events where resolving stellar populations within a galaxy is possible. At higher redshifts, where the majority of SNe will be discovered, such observations are not feasible. This, in effect, would lead to more contamination from these events in the population used for cosmology. 

As separating SNe~Ia by their global and/or local stellar population properties is extremely difficult, if not impossible, and obtaining a spectrum for the majority of SNe discovered by the VRO is too ambitious, a more concerted effort needs to be placed on distinguishing SNe~Ia based on their photometric properties. Indeed, most current methods using SNe~Ia as probes for cosmology impose a color cut using SALT2 of $|c| \lesssim 0.3$ (similar to a cut in $B-V$; \citealt{sds2020, pbb+2021}). As strong Na~I~D lines imply redder SN, either due to actual extinction or association with the intrinsically redder sub-group of high velocity SNe~Ia, we can use this knowledge to make more informed color cuts such that high velocity SNe~Ia are not included in the cosmology sample. Furthermore, if future analyses find that the "blueshifted" Na~I~D subset of SNe~Ia include both the high and lower velocity SNe~Ia that follow the \citet{pnk2019} Si~II velocity versus peak brightness relationship, then slightly more complicated cuts will be required. The upcoming release of the ZTF DR2 SNe~Ia sample, comprised of over 3000 SNe, will greatly improve our understanding of these relationships \citep{Rigault2024}.


\section*{Acknowledgements}
We thank Adam Miller and Wen-fai Fong for valuable comments and suggestions. A.E.N. acknowledges support from the Henry Luce Foundation through a Graduate Fellowship in Physics and Astronomy. The Fong Group at Northwestern acknowledges support by the National Science Foundation under grant Nos. AST-1814782, AST-1909358 and CAREER grant No. AST-2047919. Funding for this research came from the Director, Office of Science, Office of High Energy Physics of the U.S. Department of Energy under Contract no. DE-AC02-05CH1123. The National Energy Research Scientific Computing Center, a DOE Advanced Scientific Computing User Facility under the same contract, provided staff, computational resources, and data storage for this project.

This research has made use of the NASA/IPAC Extragalactic Database, which is funded by the National Aeronautics and Space Administration and operated by the California Institute of Technology.

This research was supported in part through the computational resources and staff contributions provided for the Quest high performance computing facility at Northwestern University, which is jointly supported by the Office of the Provost, the Office for Research, and Northwestern University Information Technology.

\bibliographystyle{aa}
\bibliography{refs}

\begin{appendix}
\onecolumn

\section{Supernova Properties} \label{appendix}
Here, we record all properties of the SNe~Ia studied in this work.

\footnotesize{
\begin{longtable}{l|cccccccccccc}
\caption{\label{tab:sndata} Supernova Data}\\
\hline \hline 
SN Name & 
Type &
$z$ & 
$\mu$ &
$M_B$ &
$M_{\sigma,B}$ & 
$B-V$ & 
$v_{SiII}$ & 
$\sigma{v_{SiII}}$& 
Offset [kpc] &
EW Na I D [\AA] & 
{$\sigma_{Na I D}$} [\AA] & 
Ref. \\
\hline\hline
\endfirsthead
\caption{continued.} \\
\hline \hline
SN Name & 
Type &
$z$ & 
$\mu$ &
$M_B$ &
$M_{\sigma,B}$ & 
$B-V$ & 
$v_{SiII}$ & 
$\sigma{v_{SiII}}$& 
Offset [kpc] &
EW Na I D [\AA] & 
{$\sigma_{Na I D}$} [\AA] & 
Ref. \\
\hline
\endhead
\hline
\endfoot
1998dh & low & 0.009 & 33.03 & -19.14 & 0.22 & 0.08 & 11.1 & 0.5 & 10.16 & 0.36 & 0.12 & 1 \\ 
1998dm & low & 0.007 & 32.25 & -17.61 & 0.28 & 0.26 & 10.6 & 0.3 & 4.86 & 0.62 & 0.13 & 1 \\ 
1999cp & low & 0.009 & 33.38 & -19.43 & 0.28 & -0.001 & 10.6 & 0.3 & 12.37 & 0.0 & 0.08 & 1 \\ 
1999dq & low & 0.014 & 33.94 & -19.47 & 0.15 & 0.11 & 10.9 & 0.1 & 2.17 & 1.0 & 0.07 & 1 \\ 
1999gp & low & 0.027 & 35.32 & -19.34 & 0.09 & -0.01 & 11.0 & 0.2 & 7.77 & 0.27 & 0.2 & 1 \\ 
2000cx & low & 0.008 & 32.69 & -19.57 & 0.24 & 0.12 & 11.7 & 0.2 & 16.21 & 0.0 & 0.2 & 1 \\ 
2000dn & low & 0.032 & 35.74 & -19.12 & 0.07 & -0.01 & 10.2 & 0.2 & 18.52 & 0.0 & 0.2 & 1 \\ 
2000dr & low & 0.019 & 34.55 & -18.57 & 0.11 & 0.12 & 10.5 & 0.3 & 7.93 & 0.0$^\dagger$ & 0.02 & 1 \\ 
2000fa & low & 0.021 & 34.9 & -19.01 & 0.10 & 0.06 & 12.0 & 0.2 & 4.01 & 0.97 & 0.19 & 1 \\ 
2001en & high & 0.016 & 34.19 & -19.12 & 0.15 & 0.01 & 12.5 & 0.4 & 8.48 & 0.98 & 0.14 & 1 \\ 
2001ep & low & 0.013 & 33.79 & -18.92 & 0.15 & 0.02 & 10.3 & 0.3 & 4.83 & 0.73 & 0.2 & 1 \\ 
2002bo & high & 0.004 & 31.99 & -18.05 & 0.29 & 0.39 & 13.0 & 0.3 & 2.1 & 2.44 & 0.09 & 1 \\ 
2002cr & low & 0.009 & 33.38 & -19.20 & 0.18 & -0.03 & 10.7 & 0.3 & 13.58 & 0.51 & 0.12 & 1 \\ 
2002dj & high & 0.009 & 33.08 & -19.11 & 0.21 & 0.12 & 14.5 & 0.3 & 2.22 & 0.8 & 0.13 & 1 \\ 
2002dl & low & 0.016 & 34.32 & -18.47 & 0.12 & 0.14 & 10.6 & 0.4 & 4.43 & 0.54 & 0.17 & 1 \\ 
2002eb & low & 0.027 & 35.43 & -19.42 & 0.08 & -0.07 & 10.3 & 0.4 & 10.72 & 0.0 & 0.2 & 1 \\ 
2002er & low & 0.009 & 33.29 & -19.03 & 0.21 & 0.14 & 11.8 & 0.4 & 2.51 & 1.75 & 0.09 & 1 \\ 
2002fk & low & 0.007 & 32.54 & -19.37 & 0.24 & -0.09 & 9.8 & 0.3 & 1.84 & 0.07 & 0.08 & 2 \\ 
2002ha & low & 0.014 & 34.09 & -19.37 & 0.14 & -0.08 & 10.8 & 0.1 & 8.71 & 0.83 & 0.16 & 1 \\ 
2002he & high & 0.025 & 35.23 & -18.98 & 0.09 & 0.02 & 12.4 & 0.1 & 21.17 & 0.0 & 0.1 & 1 \\ 
2003cg & low & 0.004 & 31.65 & -15.81 & 0.32 & 1.22 & 10.9 & 0.2 & 1.6 & 5.43 & 0.23 & 1 \\ 
2003fa & low & 0.040 & 36.33 & -19.72 & 0.07 & -0.09 & 11.2 & 0.4 & 39.33 & 0.0 & 0.0 & 1 \\ 
2003gn & low & 0.034 & 35.90 & -18.55 & 0.11 & 0.04 & 12.0 & 0.3 & 13.32 & 0.0$^\dagger$ & 0.01 & 1 \\ 
2003gt & low & 0.015 & 34.26 & -19.29 & 0.13 & 0.05 & 11.1 & 0.4 & 3.06 & 0.0$^\dagger$ & 0.08 & 1 \\ 
2004at & low & 0.022 & 35.08 & -19.40 & 0.09 & -0.11 & 11.3 & 0.1 & 42.3 & 0.19 & 0.14 & 1 \\ 
2004dt & high & 0.020 & 34.64 & -19.41 & 0.10 & -0.004 & 13.3 & 0.3 & 45.09 & 0.46 & 0.21 & 1 \\ 
2004ef & high & 0.031 & 35.67 & -18.80 & 0.10 & 0.08 & 12.8 & 0.4 & 6.69 & 0.9 & 0.27 & 1 \\ 
2004eo & low & 0.016 & 34.32 & -19.22 & 0.14 & 0.02 & 10.8 & 0.4 & 18.03 & 0.0 & 0.2 & 1 \\ 
2005cf & low & 0.006 & 32.59 & -19.28 & 0.25 & 0.03 & 10.1 & 0.1 & 17.25 & 0.3 & 0.06 & 1 \\ 
2005de & low & 0.015 & 34.37 & -18.93 & 0.13 & 0.08 & 10.2 & 0.2 & 14.4 & 0.51 & 0.1 & 1 \\ 
2005ki & low & 0.019 & 34.84 & -19.30 & 0.10 & -0.01 & 11.1 & 0.2 & 29.6 & 0.3 & 0.09 & 1 \\ 
2006cp & high & 0.022 & 35.15 & -19.23 & 0.11 & -0.02 & 13.6 & 0.3 & 12.26 & 0.26 & 0.14 & 1 \\ 
2006gr & low & 0.035 & 35.92 & -18.98 & 0.09 & 0.08 & 11.4 & 0.3 & 24.13 & 1.48 & 0.23 & 1 \\ 
2006le & low & 0.017 & 34.46 & -19.50 & 0.11 & -0.04 & 11.6 & 0.3 & 14.92 & 0.1 & 0.08 & 1 \\ 
2007af & low & 0.005 & 31.77 & -18.61 & 0.28 & 0.05 & 10.6 & 0.1 & 6.01 & 0.31 & 0.08 & 2 \\ 
2007le & high & 0.007 & 32.41 & -18.54 & 0.27 & 0.27 & 14.0 & 0.4 & 2.42 & 1.52 & 0.06 & 1 \\ 
2007qe & high & 0.024 & 35.10 & -19.08 & 0.11 & 0.06 & 14.1 & 0.3 & 5.9 & 0.11 & 0.07 & 1 \\ 
2008bf & low & 0.024 & 35.33 & -19.64 & 0.07 & -0.15 & 11.5 & 0.2 & 25.73 & 0.08 & 0.15 & 1 \\ 
2008ec & low & 0.016 & 34.28 & -18.72 & 0.13 & 0.15 & 10.5 & 0.1 & 4.71 & 0.61 & 0.12 & 1 \\ 
2001V & low & 0.015 & 34.36 & -19.69 & 0.13 & 0.07 & 11.6 & 0.3 & 20.02 & 1.26 & 0.11 & 1 \\ 
2005hk & low & 0.013 & 33.75 & -17.91 & 0.16 & 0.15 & 5.7 & 0.3 & 4.52 & 0.0$^\dagger$ & 0.02 & 1 \\ 
2006ax & low & 0.017 & 34.57 & -19.54 & 0.11 & -0.03 & 10.5 & 0.1 & 20.34 & 0.0$^\dagger$ & 0.1 & 1 \\ 
2006lf & low & 0.013 & 33.83 & -19.62 & 0.16 & -0.06 & 11.4 & 0.4 & 4.67 & 0.05$^\dagger$ & 0.05 & 1 \\ 
2007bd & high & 0.026 & 35.42 & -18.84 & 0.07 & 0.01 & 12.8 & 0.4 & 5.41 & 0.45 & 0.09 & 1 \\ 
2007ci & low & 0.018 & 34.71 & -18.85 & 0.11 & 0.02 & 11.8 & 0.2 & 5.08 & 0.1 & 0.15 & 1 \\ 
2005kc & low & 0.015 & 34.15 & -18.59 & 0.16 & 0.16 & 10.6 & 0.3 & 3.05 & 1.78 & 0.13 & 1 \\ 
2007on & low & 0.006 & 31.54 & -18.49 & 0.23 & 0.11 & 11.3 & 0.3 & 8.98 & 0.02$^\dagger$ & 0.0 & 3 \\ 
2008gp & low & 0.033 & 35.77 & -19.29 & 0.08 & 0.004 & 11.3 & 0.4 & 11.74 & 0.0 & 0.14 & 1 \\ 
2008hv & low & 0.013 & 33.9 & -19.11 & 0.13 & 0.05 & 10.9 & 0.2 & 10.01 & 0.0 & 0.19 & 1 \\ 
2003W & high & 0.020 & 34.87 & -18.94 & 0.10 & 0.18 & 14.8 & 0.4 & 1.45 & 0.99 & 0.2 & 1 \\ 
2003Y & faint & 0.017 & 34.47 & -16.74 & 0.14 & 0.81 & 9.8 & 0.3 & 7.03 & 0.0$^\dagger$ & 0.08 & 1 \\ 
2005M & low & 0.025 & 35.28 & -19.39 & 0.09 & -0.05 & 10.6 & 0.2 & 3.35 & 0.25 & 0.13 & 1 \\ 
2006X & high & 0.005 & 31.04 & -15.64 & 0.30 & 1.35 & 14.7 & 0.1 & 6.26 & 2.05 & 0.08 & 3 \\ 
1991bg$^*$ & faint & 0.003 & 31.40 & -16.79 & 0.05 & 0.87 & 9.63 & 0.05 & 4.07 & 0.0$^\dagger$ & 0.1 & 3 \\ 
1999aa$^*$ & low & 0.014 & 34.15 & -19.57 & 0.03 & -0.08 & 10.35 & 0.05 & 8.94 & 0.05 & 0.12 & 1 \\ 
1998bu$^*$ & low & 0.003 & 30.20 & -18.17 & 0.02 & 0.32 & 10.75 & 0.2 & 3.43 & 0.45 & 0.1 & 3 \\ 
1998bp$^*$ & low & 0.010 & 33.65 & -18.65 & 0.05 & 0.16 & 10.92 & 0.2 & 2.94 & 0.0 & 0.14 & 2 \\ 
1998aq$^*$ & low & 0.004 & 31.72 & -19.46 & 0.02 & -0.13 & 10.74 & 0.2 & 1.78 & 0.16 & 0.08 & 1 \\ 
1996X$^*$ & low & 0.007 & 32.24 & -19.51 & 0.05 & -0.10 & 11.17 & 0.22 & 8.85 & 0.0 & 0.12 & 1 \\ 
1995D$^*$ & low & 0.007 & 32.61 & -19.65 & 0.06 & -0.13 & 10.17 & 0.05 & 12.5 & 0.33 & 0.09 & 1 \\ 
1994ae$^*$ & low & 0.004 & 32.12 & -19.28 & 0.06 & -0.07 & 10.98 & 0.23 & 2.74 & 0.29 & 0.06 & 2 \\ 
1990N$^*$ & low & 0.003 & 31.81 & -19.23 & 0.05 & -0.003 & 9.38 & 0.22 & 4.45 & 0.0$^\dagger$ & 0.1 & 2 \\ 
1986G$^*$ & low & 0.002 & 28.19 & -16.58 & 0.08 & 0.82 & 10.47 & 0.05 & 4.08 & 0.34 & 0.01 & 3 \\ 
2011fe$^*$ & low & 0.001 & 29.18 & -19.20 & 0.30 & -0.02 & 10.58 & 0.05 & 4.67 & 0.07 & 0.05 & 2 \\ 
2008Q$^*$ & low & 0.008 & 32.69 & -19.51 & 0.03 & -0.07 & 11.42 & 0.2 & 24.69 & 0.0$^\dagger$ & 0.04 & 2 \\ 
2006D$^*$ & low & 0.009 & 32.92 & -18.96 & 0.04 & 0.029 & 10.42 & 0.05 & 2.22 & 0.0 & 0.05 & 2 \\ 
2005ke$^*$ & faint & 0.005 & 31.57 & -16.79 & 0.05 & 0.72 & 8.97 & 0.2 & 5.76 & 0.0 & 0.1 & 1 \\ 
2003hv$^*$ & low & 0.006 & 31.89 & -19.47 & 0.03 & -0.07 & 10.91 & 0.2 & 6.56 & 0.0 & 0.1 & 1 \\ 
2003du$^*$ & low & 0.006 & 32.85 & -19.46 & 0.03 & -0.12 & 10.45 & 0.2 & 2.04 & 0.1 & 0.07 & 2 \\ 
2002dp$^*$ & low & 0.012 & 33.57 & -19.15 & 0.03 & 0.09 & 11.48 & 0.2 & 10.68 & 0.7 & 0.14 & 1 \\ 
2000E$^*$ & high & 0.005 & 32.09 & -20.58 & 0.24 & -0.15 & 19.22 & 0.5 & 2.99 & 0.0$^\dagger$ & 0.05 & 1 \\ 
1999by$^*$ & faint & 0.002 & 30.74 & -17.26 & 0.03 & 0.43 & 10.0 & 0.2 & 5.99 & 0.0 & 0.2 & 4 \\ 
1991T$^*$ & low & 0.006 & 30.76 & -19.24 & 0.02 & 0.13 & 10.0 & 0.4 & 6.12 & 1.3 & 0.14 & 5 \\ \hline
\end{longtable}}
\vspace{-0.1in}
\tablefoot{The types of each SNe~Ia studied in this work (low velocity, high velocity, or faint), the redshifts and Hubble flow distance moduli ($\mu$) of their hosts, their peak $B$-band absolute magnitude ($M_B$) and error ($M_{\sigma,B}$), Si II velocities ($v_{SiII}$) and errors ($\sigma{v_{SiII}}$), physical offsets, and Na I D equivalent widths (EW) and errors ($\sigma_{Na I D}$). The references refer to the $\mu$ measurements. All redshifts were acquired from NED. The $B-V$ colors and Si II velocities were collected from \citet{zkf2018} and \citet{pnk2021}. We correct the $M_B$ values in the \citet{zkf2018} and \citet{pnk2021} samples using our collected $\mu$. We determine the all of the physical offsets and several of the Na I D EW and errors; the rest are from \citet{wang2019}. \\ 
$^*$ \citet{pnk2021} SNe~Ia sample (the rest are from \citealt{zkf2017}) \\
$^\dagger$ Na I D EW determined in this work (all others collected in \citealt{wang2019}). \\
References: (1) \citealt{mould2000}; (2) \citealt{riess2022}; (3) \citealt{Blakeslee2010}; (4) \citealt{macri2001}; (5) \citealt{saha2001}}

\section{Host Galaxy Photometry} \label{appendix}
Here, we record all host galaxy photometry used in the \texttt{Prospector} modeling.

\footnotesize{
\begin{longtable}{l|cccc}
\caption{Photometric Catalog \label{tab:phot}} \\
\hline \hline 
SN Name	&
R.A. (J2000) &
Decl. (J2000) &
Filter &
$m_{\rm AB}$ \\
\hline\hline
\endfirsthead
\caption{continued.} \\
\hline \hline
SN Name	&
R.A. (J2000) &
Decl. (J2000) &
Filter &
$m_{\rm AB}$ \\
\hline
\endhead
\hline
\endfoot
1998dh & 348.6829 & 4.5344 & U & 10.45 $\pm$ 0.17 \\
& & & B & 12.0 $\pm$ 0.17 \\
& & & V & 11.53 $\pm$ 0.17 \\
& & & R & 11.45 $\pm$ 0.13 \\
& & & I & 10.58 $\pm$ 0.13 \\
& & & J & 10.29 $\pm$ 0.13 \\
& & & H & 10.22 $\pm$ 0.13 \\
& & & K & 10.17 $\pm$ 0.13 \\ \hline
1998dm & 21.5601 & -6.0941 & FUV & 15.86 $\pm$ 0.13 \\
& & & NUV & 15.35 $\pm$ 0.13 \\
& & & u & 14.52 $\pm$ 0.13 \\
& & & g & 13.29 $\pm$ 0.13 \\
& & & r & 12.78 $\pm$ 0.13 \\
& & & i & 12.54 $\pm$ 0.13 \\
& & & z & 12.43 $\pm$ 0.13 \\
& & & J & 12.42 $\pm$ 0.13 \\
& & & H & 12.27 $\pm$ 0.13 \\
& & & K & 12.46 $\pm$ 0.13 \\
& & & W1 & 14.77 $\pm$ 0.13 \\
& & & W2 & 13.47 $\pm$ 0.13 \\ \hline
1999cp & 211.6454 & -5.4531 & FUV & 13.78 $\pm$ 0.11 \\
& & & NUV & 13.46 $\pm$ 0.11 \\
& & & U & 13.53 $\pm$ 0.69 \\
& & & B & 12.74 $\pm$ 0.76 \\
& & & V & 12.44 $\pm$ 0.68 \\
& & & J & 12.43 $\pm$ 0.13 \\
& & & H & 12.25 $\pm$ 0.13 \\
& & & K & 12.46 $\pm$ 0.13 \\
& & & W1 & 13.19 $\pm$ 0.13 \\
& & & W2 & 13.72 $\pm$ 0.13 \\
& & & W3 & 11.84 $\pm$ 0.13 \\
& & & W4 & 11.05 $\pm$ 0.13 \\ \hline
1999dq & 38.5001 & 20.9768 & FUV & 16.34 $\pm$ 0.11 \\
& & & NUV & 15.88 $\pm$ 0.11 \\
& & & U & 13.7 $\pm$ 0.17 \\
& & & B & 12.69 $\pm$ 0.17 \\
& & & V & 12.36 $\pm$ 0.14 \\
& & & J & 11.05 $\pm$ 0.13 \\
& & & H & 10.85 $\pm$ 0.13 \\
& & & K & 10.99 $\pm$ 0.13 \\
& & & W1 & 11.84 $\pm$ 0.13 \\
& & & W2 & 12.38 $\pm$ 0.13 \\
& & & W3 & 10.84 $\pm$ 0.13 \\
& & & W4 & 10.68 $\pm$ 0.13 \\ \hline
1999gp & 37.9169 & 39.3784 & B & 13.72 $\pm$ 0.27 \\
& & & J & 12.66 $\pm$ 0.13 \\
& & & H & 12.43 $\pm$ 0.13 \\
& & & K & 12.69 $\pm$ 0.13 \\
& & & W1 & 13.44 $\pm$ 0.13 \\
& & & W2 & 14.03 $\pm$ 0.13 \\
& & & W3 & 13.18 $\pm$ 0.13 \\
& & & W4 & 13.43 $\pm$ 0.13 \\ \hline
2000cx & 21.1988 & 9.5388 & FUV & 18.43 $\pm$ 0.11 \\
& & & NUV & 15.88 $\pm$ 0.17 \\
& & & U & 12.3 $\pm$ 0.13 \\
& & & B & 10.82 $\pm$ 0.13 \\
& & & V & 10.02 $\pm$ 0.13 \\
& & & r & 10.57 $\pm$ 0.13 \\
& & & J & 9.03 $\pm$ 0.13 \\
& & & H & 8.83 $\pm$ 0.13 \\
& & & K & 9.05 $\pm$ 0.13 \\
& & & W1 & 10.64 $\pm$ 0.13 \\
& & & W2 & 11.33 $\pm$ 0.13 \\
& & & W3 & 12.02 $\pm$ 0.13 \\
& & & W4 & 12.53 $\pm$ 0.13 \\ \hline
2000dn & 346.2815 & -3.2045 & B & 14.65 $\pm$ 0.15 \\
& & & J & 12.8 $\pm$ 0.13 \\
& & & H & 12.58 $\pm$ 0.13 \\
& & & K & 12.78 $\pm$ 0.13 \\
& & & W1 & 13.93 $\pm$ 0.13 \\
& & & W2 & 14.67 $\pm$ 0.13 \\
& & & W3 & 14.81 $\pm$ 0.13 \\
& & & W4 & 15.15 $\pm$ 0.9 \\ \hline
2000dr & 15.4274 & -15.5678 & FUV & 19.95 $\pm$ 0.26 \\
& & & NUV & 18.32 $\pm$ 0.11 \\
& & & B & 14.32 $\pm$ 0.15 \\
& & & J & 11.82 $\pm$ 0.13 \\
& & & H & 11.63 $\pm$ 0.13 \\
& & & K & 11.82 $\pm$ 0.13 \\
& & & W1 & 13.16 $\pm$ 0.13 \\
& & & W2 & 13.9 $\pm$ 0.13 \\
& & & W3 & 14.76 $\pm$ 0.13 \\
& & & W4 & 15.38 $\pm$ 1.5 \\ \hline
2000fa & 108.8732 & 23.4262 & B & 14.48 $\pm$ 0.69 \\
& & & J & 13.58 $\pm$ 0.13 \\
& & & H & 13.56 $\pm$ 0.13 \\
& & & K & 13.22 $\pm$ 0.13 \\ \hline
2001en & 21.3364 & 34.025 & FUV & 19.2 $\pm$ 0.11 \\
& & & NUV & 18.2 $\pm$ 0.11 \\
& & & U & 14.22 $\pm$ 1.83 \\
& & & B & 13.13 $\pm$ 1.74 \\
& & & V & 12.56 $\pm$ 1.79 \\
& & & J & 11.68 $\pm$ 0.13 \\
& & & H & 11.46 $\pm$ 0.13 \\
& & & K & 11.61 $\pm$ 0.13 \\
& & & W1 & 13.9 $\pm$ 0.13 \\
& & & W2 & 14.43 $\pm$ 0.13 \\
& & & W3 & 13.15 $\pm$ 0.13 \\
& & & W4 & 12.25 $\pm$ 0.13 \\ \hline
2001ep & 74.2486 & -4.7569 & J & 12.58 $\pm$ 0.13 \\
& & & H & 12.43 $\pm$ 0.13 \\
& & & K & 12.61 $\pm$ 0.13 \\
& & & W1 & 14.4 $\pm$ 0.13 \\
& & & W2 & 15.05 $\pm$ 0.13 \\
& & & W3 & 13.82 $\pm$ 0.13 \\
& & & W4 & 13.07 $\pm$ 0.13 \\ \hline
2002bo & 154.5235 & 21.8323 & FUV & 17.67 $\pm$ 0.11 \\
& & & NUV & 16.08 $\pm$ 0.11 \\
& & & u & 13.19 $\pm$ 0.13 \\
& & & B & 11.85 $\pm$ 0.13 \\
& & & g & 11.54 $\pm$ 0.13 \\
& & & V & 11.08 $\pm$ 0.13 \\
& & & r & 10.74 $\pm$ 0.13 \\
& & & i & 10.3 $\pm$ 0.13 \\
& & & z & 9.99 $\pm$ 2.09 \\
& & & J & 9.52 $\pm$ 0.13 \\
& & & H & 9.25 $\pm$ 0.13 \\
& & & K & 9.43 $\pm$ 0.13 \\
& & & W1 & 11.23 $\pm$ 0.13 \\
& & & W2 & 11.82 $\pm$ 0.13 \\
& & & W3 & 11.44 $\pm$ 0.13 \\
& & & W4 & 11.38 $\pm$ 0.13 \\ \hline
2002cr & 211.6454 & -5.4531 & FUV & 13.66 $\pm$ 0.13 \\
& & & NUV & 13.36 $\pm$ 0.13 \\
& & & U & 13.53 $\pm$ 0.69 \\
& & & B & 12.74 $\pm$ 0.68 \\
& & & V & 12.44 $\pm$ 0.68 \\
& & & J & 12.43 $\pm$ 0.13 \\
& & & H & 12.25 $\pm$ 0.13 \\
& & & K & 12.46 $\pm$ 0.13 \\
& & & W1 & 14.18 $\pm$ 0.13 \\
& & & W2 & 14.67 $\pm$ 0.13 \\
& & & W3 & 12.27 $\pm$ 0.13 \\
& & & W4 & 11.05 $\pm$ 0.13 \\ \hline
2002dj & 198.2543 & -19.5182 & FUV & 18.46 $\pm$ 0.13 \\
& & & NUV & 16.18 $\pm$ 0.13 \\
& & & U & 12.51 $\pm$ 0.16 \\
& & & V & 10.51 $\pm$ 0.17 \\
& & & R & 10.29 $\pm$ 0.13 \\
& & & J & 9.56 $\pm$ 0.13 \\
& & & H & 9.41 $\pm$ 0.13 \\
& & & K & 9.59 $\pm$ 0.13 \\
& & & W1 & 11.38 $\pm$ 0.13 \\
& & & W2 & 11.97 $\pm$ 0.13 \\
& & & W3 & 12.51 $\pm$ 0.13 \\
& & & W4 & 12.1 $\pm$ 0.13 \\ \hline
2002dl & 335.2213 & 33.2953 & B & 14.4 $\pm$ 0.27 \\
& & & I & 12.74 $\pm$ 0.13 \\
& & & J & 12.18 $\pm$ 0.13 \\
& & & H & 11.88 $\pm$ 0.13 \\
& & & K & 12.02 $\pm$ 0.13 \\
& & & W1 & 13.92 $\pm$ 0.13 \\
& & & W2 & 14.45 $\pm$ 0.13 \\
& & & W3 & 13.42 $\pm$ 0.13 \\
& & & W4 & 12.98 $\pm$ 0.13 \\ \hline
2002eb & 334.7762 & 24.5982 & FUV & 18.25 $\pm$ 0.13 \\
& & & NUV & 17.9 $\pm$ 0.13 \\
& & & B & 14.91 $\pm$ 0.27 \\
& & & J & 13.54 $\pm$ 0.13 \\
& & & H & 14.78 $\pm$ 0.13 \\
& & & K & 14.81 $\pm$ 0.13 \\
& & & W1 & 14.57 $\pm$ 0.13 \\
& & & W2 & 14.7 $\pm$ 0.13 \\
& & & W3 & 13.0 $\pm$ 0.13 \\
& & & W4 & 11.84 $\pm$ 0.13 \\ \hline
2002er & 247.878 & 7.9946 & FUV & 19.86 $\pm$ 0.31 \\
& & & NUV & 18.5 $\pm$ 0.13 \\
& & & B & 14.43 $\pm$ 0.14 \\
& & & V & 13.94 $\pm$ 0.14 \\
& & & J & 12.3 $\pm$ 0.13 \\
& & & H & 12.79 $\pm$ 0.13 \\
& & & K & 12.31 $\pm$ 0.13 \\
& & & W1 & 13.36 $\pm$ 0.13 \\
& & & W2 & 13.75 $\pm$ 0.13 \\
& & & W3 & 11.4 $\pm$ 0.13 \\
& & & W4 & 10.26 $\pm$ 0.13 \\ \hline
2002fk & 50.5273 & -15.4001 & FUV & 15.6 $\pm$ 0.13 \\
& & & NUV & 14.79 $\pm$ 0.13 \\
& & & U & 12.38 $\pm$ 0.17 \\
& & & B & 11.65 $\pm$ 0.17 \\
& & & V & 11.42 $\pm$ 0.17 \\
& & & J & 10.98 $\pm$ 0.13 \\
& & & H & 10.83 $\pm$ 0.13 \\
& & & K & 11.05 $\pm$ 0.13 \\
& & & W1 & 13.42 $\pm$ 0.13 \\
& & & W2 & 14.01 $\pm$ 0.13 \\
& & & W3 & 13.16 $\pm$ 0.13 \\
& & & W4 & 12.47 $\pm$ 0.13 \\ \hline
2002ha & 311.8294 & 0.3208 & FUV & 17.75 $\pm$ 0.13 \\
& & & NUV & 16.81 $\pm$ 0.13 \\
& & & U & 13.69 $\pm$ 0.39 \\
& & & B & 12.46 $\pm$ 0.31 \\
& & & V & 11.86 $\pm$ 0.59 \\
& & & R & 11.3 $\pm$ 0.13 \\
& & & I & 10.85 $\pm$ 0.13 \\
& & & J & 10.73 $\pm$ 0.13 \\
& & & H & 10.56 $\pm$ 0.13 \\
& & & K & 10.73 $\pm$ 0.13 \\
& & & W1 & 12.67 $\pm$ 0.13 \\
& & & W2 & 13.37 $\pm$ 0.13 \\
& & & W3 & 14.15 $\pm$ 0.13 \\
& & & W4 & 14.36 $\pm$ 0.31 \\ \hline
2002he & 125.0072 & 62.8306 & J & 12.8 $\pm$ 0.13 \\
& & & W2 & 13.62 $\pm$ 0.13 \\
& & & W3 & 15.35 $\pm$ 0.13 \\ \hline
2003cg & 153.5627 & 3.4661 & FUV & 16.78 $\pm$ 0.13 \\
& & & NUV & 15.81 $\pm$ 0.13 \\
& & & U & 11.96 $\pm$ 0.18 \\
& & & B & 10.79 $\pm$ 0.13 \\
& & & V & 10.14 $\pm$ 0.13 \\
& & & I & 9.64 $\pm$ 0.14 \\
& & & J & 11.26 $\pm$ 0.13 \\
& & & H & 9.02 $\pm$ 0.13 \\
& & & K & 9.21 $\pm$ 0.13 \\
& & & W1 & 11.07 $\pm$ 0.13 \\
& & & W2 & 11.56 $\pm$ 0.13 \\
& & & W3 & 10.54 $\pm$ 0.13 \\
& & & W4 & 10.15 $\pm$ 0.13 \\ \hline
2003fa & 266.0358 & 40.8573 & B & 15.24 $\pm$ 0.54 \\
& & & J & 13.64 $\pm$ 0.13 \\
& & & H & 13.43 $\pm$ 1.45 \\
& & & K & 13.51 $\pm$ 0.13 \\
& & & W1 & 14.55 $\pm$ 0.13 \\
& & & W2 & 15.03 $\pm$ 0.13 \\
& & & W3 & 13.0 $\pm$ 0.22 \\
& & & W4 & 12.16 $\pm$ 0.13 \\ \hline
2003gn & 338.4603 & 20.8053 & FUV & 18.63 $\pm$ nan \\
& & & NUV & 18.07 $\pm$ 0.13 \\
& & & B & 14.89 $\pm$ 0.27 \\
& & & J & 16.0 $\pm$ 0.15 \\
& & & H & 13.63 $\pm$ 0.13 \\
& & & K & 15.85 $\pm$ 0.13 \\
& & & W1 & 15.12 $\pm$ 0.13 \\
& & & W2 & 15.72 $\pm$ 0.13 \\
& & & W3 & 14.25 $\pm$ 0.59 \\
& & & W4 & 13.78 $\pm$ 0.13 \\ \hline
2003gt & 308.245 & 9.8744 & FUV & 18.24 $\pm$ 0.15 \\
& & & NUV & 17.57 $\pm$ 0.13 \\
& & & U & 14.08 $\pm$ 0.26 \\
& & & B & 13.05 $\pm$ 0.23 \\
& & & V & 12.46 $\pm$ 0.24 \\
& & & J & 13.58 $\pm$ 0.13 \\
& & & H & 13.26 $\pm$ 0.13 \\
& & & K & 13.41 $\pm$ 0.13 \\
& & & W1 & 13.71 $\pm$ 0.13 \\
& & & W2 & 14.3 $\pm$ 0.13 \\
& & & W3 & 13.01 $\pm$ 0.13 \\
& & & W4 & 12.09 $\pm$ 0.13 \\ \hline
2004at & 164.6954 & 59.4868 & u & 16.88 $\pm$ 0.13 \\
& & & g & 15.87 $\pm$ 0.13 \\
& & & r & 15.48 $\pm$ 0.13 \\
& & & i & 15.31 $\pm$ 0.13 \\
& & & z & 15.12 $\pm$ 0.13 \\
& & & J & 16.39 $\pm$ 0.18 \\
& & & K & 16.59 $\pm$ 0.16 \\ \hline
2004dt & 30.5514 & -0.1006 & FUV & 19.37 $\pm$ 0.13 \\
& & & NUV & 18.31 $\pm$ 0.13 \\
& & & u & 15.37 $\pm$ 0.13 \\
& & & B & 13.83 $\pm$ 0.64 \\
& & & g & 13.71 $\pm$ 0.13 \\
& & & V & 12.89 $\pm$ 0.25 \\
& & & r & 12.95 $\pm$ 0.13 \\
& & & i & 12.57 $\pm$ 0.13 \\
& & & z & 12.33 $\pm$ 0.13 \\
& & & J & 12.25 $\pm$ 0.13 \\
& & & H & 12.07 $\pm$ 0.13 \\
& & & K & 12.26 $\pm$ 0.13 \\
& & & W1 & 13.75 $\pm$ 0.13 \\
& & & W2 & 14.47 $\pm$ 0.13 \\
& & & W3 & 14.29 $\pm$ 0.13 \\
& & & W4 & 14.14 $\pm$ 0.16 \\ \hline
2004ef & 340.5438 & 19.997 & NUV & 18.5 $\pm$ 0.13 \\
& & & B & 14.38 $\pm$ 0.5 \\
& & & J & 14.8 $\pm$ 0.13 \\
& & & H & 14.67 $\pm$ 0.13 \\
& & & K & 14.94 $\pm$ 0.13 \\
& & & W1 & 14.82 $\pm$ 0.13 \\
& & & W2 & 15.57 $\pm$ 0.13 \\ 
& & & W3 & 16.31 $\pm$ 0.24 \\ \hline
2004eo & 308.2092 & 9.9264 & U & 13.79 $\pm$ 0.22 \\
& & & B & 12.59 $\pm$ 0.21 \\
& & & V & 11.92 $\pm$ 0.21 \\
& & & J & 10.72 $\pm$ 0.13 \\
& & & H & 10.47 $\pm$ 0.13 \\
& & & K & 10.6 $\pm$ 0.13 \\
& & & W1 & 12.24 $\pm$ 0.13 \\
& & & W2 & 12.82 $\pm$ 0.13 \\
& & & W3 & 12.18 $\pm$ 0.13 \\
& & & W4 & 11.86 $\pm$ 0.13 \\ \hline
2005cf & 230.3857 & -7.3772 & FUV & 14.66 $\pm$ 0.13 \\
& & & NUV & 14.39 $\pm$ 0.13 \\
& & & B & 13.26 $\pm$ 2.11 \\
& & & R & 13.67 $\pm$ nan \\
& & & I & 13.7 $\pm$ 4.31 \\
& & & J & 12.56 $\pm$ 0.13 \\
& & & H & 12.39 $\pm$ 0.13 \\
& & & K & 12.71 $\pm$ 0.13 \\
& & & W1 & 14.07 $\pm$ 0.13 \\
& & & W2 & 14.56 $\pm$ 0.13 \\
& & & W3 & 12.19 $\pm$ 0.13 \\
& & & W4 & 10.56 $\pm$ 0.13 \\ \hline
2005de & 270.6036 & 36.0404 & B & 14.48 $\pm$ 0.29 \\
& & & J & 12.44 $\pm$ 0.13 \\
& & & H & 12.22 $\pm$ 0.13 \\
& & & K & 12.35 $\pm$ 0.13 \\ \hline
2005ki & 160.1182 & 9.1826 & u & 15.15 $\pm$ 0.13 \\
& & & g & 13.27 $\pm$ 0.13 \\
& & & r & 12.47 $\pm$ 0.13 \\
& & & i & 12.06 $\pm$ 0.13 \\
& & & z & 11.81 $\pm$ 0.13 \\
& & & J & 11.4 $\pm$ 0.13 \\
& & & H & 11.23 $\pm$ 0.13 \\
& & & K & 11.39 $\pm$ 0.13 \\
& & & W1 & 13.07 $\pm$ 0.13 \\
& & & W2 & 13.75 $\pm$ 0.13 \\ \hline
2006cp & 184.8058 & 22.4315 & u & 15.45 $\pm$ 0.13 \\
& & & g & 14.33 $\pm$ 0.13 \\
& & & r & 13.84 $\pm$ 0.13 \\
& & & i & 13.62 $\pm$ 0.13 \\
& & & z & 13.5 $\pm$ 0.13 \\
& & & J & 13.13 $\pm$ 0.17 \\
& & & H & 12.38 $\pm$ 0.21 \\
& & & K & 12.12 $\pm$ 0.28 \\ \hline
2006gr & 338.1021 & 30.8356 & B & 14.49 $\pm$ 0.27 \\
& & & J & 14.03 $\pm$ 0.13 \\
& & & H & 13.76 $\pm$ 0.13 \\
& & & K & 13.94 $\pm$ 0.13 \\
& & & W1 & 14.1 $\pm$ 0.13 \\
& & & W2 & 14.66 $\pm$ 0.13 \\
& & & W3 & 12.95 $\pm$ 0.13 \\
& & & W4 & 12.5 $\pm$ 0.13 \\ \hline
2006le & 75.1821 & 62.244 & B & 12.64 $\pm$ 0.32 \\
& & & J & 11.07 $\pm$ 0.13 \\
& & & H & 10.85 $\pm$ 0.13 \\
& & & K & 11.07 $\pm$ 0.13 \\
& & & W1 & 13.05 $\pm$ 0.13 \\
& & & W2 & 13.61 $\pm$ 0.13 \\
& & & W3 & 11.99 $\pm$ 0.13 \\
& & & W4 & 11.51 $\pm$ 0.13 \\ \hline
2007af & 215.599 & -0.3877 & NUV & 15.12 $\pm$ 0.13 \\
& & & u & 14.71 $\pm$ 0.13 \\
& & & B & 14.25 $\pm$ 0.13 \\
& & & g & 13.64 $\pm$ 0.13 \\
& & & r & 13.02 $\pm$ 0.13 \\
& & & i & 12.65 $\pm$ 0.13 \\
& & & z & 12.57 $\pm$ 0.13 \\ \hline
2007le & 354.7027 & -6.5179 & FUV & 16.7 $\pm$ 0.13 \\
& & & NUV & 15.86 $\pm$ 0.13 \\
& & & U & 12.68 $\pm$ 0.21 \\
& & & B & 11.93 $\pm$ 0.21 \\
& & & V & 11.51 $\pm$ 0.21 \\
& & & J & 10.66 $\pm$ 0.13 \\
& & & H & 10.49 $\pm$ 0.13 \\
& & & K & 10.7 $\pm$ 0.13 \\
& & & W1 & 13.29 $\pm$ 0.13 \\
& & & W2 & 13.85 $\pm$ 0.28 \\
& & & W3 & 11.95 $\pm$ 0.13 \\
& & & W4 & 11.27 $\pm$ 0.13 \\ \hline
2007qe & 358.5504 & 27.409 & FUV & 20.16 $\pm$ 0.28 \\
& & & NUV & 21.14 $\pm$ 0.13 \\
& & & u & 18.52 $\pm$ 0.13 \\
& & & g & 17.39 $\pm$ 0.13 \\
& & & r & 16.92 $\pm$ 0.13 \\
& & & i & 16.68 $\pm$ 0.13 \\
& & & z & 16.72 $\pm$ 0.13 \\
& & & J & 17.62 $\pm$ 0.29 \\
& & & W1 & 17.76 $\pm$ 0.13 \\
& & & W2 & 19.03 $\pm$ 0.77 \\
& & & W3 & 14.59 $\pm$ 0.13 \\ \hline
2008bf & 181.0068 & 20.2724 & B & 13.57 $\pm$ 0.21 \\
& & & g & 13.73 $\pm$ 0.13 \\
& & & V & 13.0 $\pm$ 0.21 \\
& & & r & 12.92 $\pm$ 0.13 \\
& & & i & 12.5 $\pm$ 0.13 \\
& & & z & 12.17 $\pm$ 0.13 \\
& & & J & 12.79 $\pm$ 0.13 \\
& & & H & 12.58 $\pm$ 0.13 \\
& & & K & 12.72 $\pm$ 0.13 \\
& & & W1 & 13.08 $\pm$ 0.13 \\
& & & W2 & 13.76 $\pm$ 0.13 \\
& & & W3 & 13.85 $\pm$ 0.13 \\
& & & W4 & 13.53 $\pm$ 0.14 \\ \hline
2008ec & 345.8151 & 8.874 & FUV & 16.25 $\pm$ 0.13 \\
& & & NUV & 15.47 $\pm$ 0.13 \\
& & & U & 13.06 $\pm$ 0.32 \\
& & & B & 12.58 $\pm$ 0.32 \\
& & & V & 12.15 $\pm$ 0.33 \\
& & & J & 11.14 $\pm$ 0.13 \\
& & & H & 10.83 $\pm$ 0.13 \\
& & & K & 10.83 $\pm$ 0.13 \\
& & & W1 & 11.1 $\pm$ 0.13 \\
& & & W2 & 10.95 $\pm$ 0.13 \\
& & & W3 & 8.88 $\pm$ 0.13 \\ \hline
2001V & 179.359 & 25.2331 & FUV & 18.18 $\pm$ 0.13 \\
& & & NUV & 17.73 $\pm$ 0.13 \\
& & & u & 16.74 $\pm$ 0.13 \\
& & & B & 15.47 $\pm$ 0.26 \\
& & & g & 15.56 $\pm$ 0.13 \\
& & & r & 15.05 $\pm$ 0.13 \\
& & & i & 14.81 $\pm$ 0.13 \\
& & & z & 14.68 $\pm$ 0.13 \\
& & & J & 14.93 $\pm$ 0.13 \\
& & & H & 14.8 $\pm$ 0.13 \\
& & & K & 15.05 $\pm$ 0.13 \\
& & & W1 & 16.11 $\pm$ 0.13 \\
& & & W2 & 16.56 $\pm$ 0.13 \\
& & & W3 & 14.84 $\pm$ 0.13 \\
& & & W4 & 14.11 $\pm$ 0.19 \\ \hline
2005hk & 6.9572 & -1.1999 & FUV & 16.48 $\pm$ 0.13 \\
& & & NUV & 16.26 $\pm$ 0.13 \\
& & & u & 15.91 $\pm$ 0.13 \\
& & & B & 14.77 $\pm$ 0.4 \\
& & & g & 14.8 $\pm$ 0.13 \\
& & & r & 14.35 $\pm$ 0.13 \\
& & & i & 14.14 $\pm$ 0.13 \\
& & & z & 14.16 $\pm$ 0.13 \\
& & & J & 15.07 $\pm$ 0.13 \\
& & & H & 14.81 $\pm$ 0.13 \\
& & & K & 15.29 $\pm$ 0.25 \\
& & & W1 & 16.49 $\pm$ 0.13 \\
& & & W2 & 17.08 $\pm$ 0.13 \\
& & & W3 & 15.88 $\pm$ 0.13 \\
& & & W4 & 14.88 $\pm$ 0.46 \\ \hline
2006ax & 170.999 & -12.2964 & r & 15.19 $\pm$ 0.13 \\
& & & W1 & 14.73 $\pm$ 0.13 \\
& & & W2 & 15.47 $\pm$ 0.13 \\
& & & W3 & 15.48 $\pm$ 0.13 \\
& & & W4 & 14.27 $\pm$ 0.27 \\ \hline
2006lf & 69.618 & 44.0371 & I & 11.2 $\pm$ 0.13 \\
& & & J & 11.23 $\pm$ 0.13 \\
& & & H & 11.05 $\pm$ 0.13 \\
& & & K & 11.27 $\pm$ 0.13 \\
& & & W1 & 13.53 $\pm$ 0.13 \\
& & & W2 & 14.09 $\pm$ 0.13 \\
& & & W3 & 13.03 $\pm$ 0.13 \\
& & & W4 & 11.9 $\pm$ 0.13 \\ \hline
2007bd & 127.887 & -1.1977 & B & 14.3 $\pm$ 0.27 \\
& & & J & 13.11 $\pm$ 0.13 \\
& & & H & 12.92 $\pm$ 0.13 \\
& & & K & 13.07 $\pm$ 0.13 \\
& & & W1 & 14.77 $\pm$ 0.13 \\
& & & W2 & 15.53 $\pm$ 0.13 \\
& & & W3 & 15.76 $\pm$ 0.13 \\
& & & W4 & 14.52 $\pm$ 0.36 \\ \hline
2007ci & 176.4421 & 19.7739 & FUV & 19.83 $\pm$ 0.4 \\
& & & NUV & 18.62 $\pm$ 0.13 \\
& & & U & 14.9 $\pm$ 0.27 \\
& & & B & 13.58 $\pm$ 0.21 \\
& & & V & 12.78 $\pm$ 0.21 \\
& & & R & 13.27 $\pm$ 0.13 \\
& & & J & 11.79 $\pm$ 0.13 \\
& & & H & 11.57 $\pm$ 0.13 \\
& & & K & 11.82 $\pm$ 0.13 \\
& & & W1 & 13.3 $\pm$ 0.13 \\
& & & W2 & 14.0 $\pm$ 0.13 \\
& & & W3 & 15.37 $\pm$ 0.13 \\
& & & W4 & 15.15 $\pm$ 0.7 \\ \hline
2005kc & 338.528 & 5.5699 & NUV & 17.07 $\pm$ 0.13 \\
& & & B & 12.71 $\pm$ 0.29 \\
& & & V & 12.42 $\pm$ 0.14 \\
& & & J & 10.82 $\pm$ 0.13 \\
& & & H & 10.6 $\pm$ 0.13 \\
& & & K & 10.8 $\pm$ 0.13 \\ \hline
2007on & 54.7145 & -35.5939 & FUV & 16.65 $\pm$ 0.13 \\
& & & NUV & 15.65 $\pm$ 0.13 \\
& & & U & 12.24 $\pm$ 0.17 \\
& & & B & 10.76 $\pm$ 0.17 \\
& & & V & 9.97 $\pm$ 0.17 \\
& & & r & 10.36 $\pm$ 0.13 \\
& & & J & 8.78 $\pm$ 0.13 \\
& & & H & 8.57 $\pm$ 0.13 \\
& & & K & 8.77 $\pm$ 0.13 \\
& & & W1 & 10.71 $\pm$ 0.13 \\
& & & W2 & 11.4 $\pm$ 0.13 \\
& & & W3 & 12.11 $\pm$ 0.13 \\
& & & W4 & 13.06 $\pm$ 0.13 \\ \hline
2008gp & 50.762 & 1.371 & B & 13.97 $\pm$ 0.32 \\
& & & J & 14.23 $\pm$ 0.13 \\
& & & H & 14.16 $\pm$ 0.13 \\
& & & K & 14.5 $\pm$ 0.13 \\
& & & W1 & 14.47 $\pm$ 0.13 \\
& & & W2 & 15.1 $\pm$ 0.13 \\
& & & W3 & 13.96 $\pm$ 0.13 \\
& & & W4 & 13.3 $\pm$ 0.13 \\ \hline
2008hv & 136.9027 & 3.3929 & B & 12.79 $\pm$ 0.21 \\
& & & g & 12.87 $\pm$ 0.13 \\
& & & r & 12.21 $\pm$ 0.13 \\
& & & I & 11.83 $\pm$ 0.13 \\
& & & z & 11.99 $\pm$ 0.13 \\
& & & J & 11.13 $\pm$ 0.13 \\
& & & H & 10.95 $\pm$ 0.13 \\
& & & K & 11.15 $\pm$ 0.13 \\
& & & W1 & 13.02 $\pm$ 0.13 \\
& & & W2 & 13.71 $\pm$ 0.13 \\
& & & W3 & 14.94 $\pm$ 0.13 \\
& & & W4 & 15.25 $\pm$ 0.66 \\ \hline
2003W & 146.7063 & 16.0429 & NUV & 17.17 $\pm$ 0.13 \\
& & & u & 12.83 $\pm$ 0.13 \\
& & & B & 13.85 $\pm$ 0.32 \\
& & & g & 13.77 $\pm$ 0.13 \\
& & & r & 13.13 $\pm$ 0.13 \\
& & & i & 12.72 $\pm$ 0.13 \\
& & & z & 12.52 $\pm$ 0.13 \\
& & & J & 9.65 $\pm$ 0.13 \\
& & & H & 12.17 $\pm$ 0.13 \\
& & & K & 12.32 $\pm$ 0.13 \\ \hline
2003Y & 133.6455 & 57.1667 & g & 13.6 $\pm$ 0.13 \\
& & & r & 12.82 $\pm$ 0.13 \\
& & & i & 12.41 $\pm$ 0.13 \\
& & & z & 12.12 $\pm$ 0.13 \\
& & & J & 11.94 $\pm$ 0.13 \\
& & & H & 11.72 $\pm$ 0.13 \\
& & & K & 11.91 $\pm$ 0.13 \\
& & & W1 & 13.36 $\pm$ 0.13 \\
& & & W2 & 14.06 $\pm$ 0.13 \\
& & & W3 & 15.06 $\pm$ 0.13 \\ \hline
2005M & 144.3858 & 23.025 & u & 15.67 $\pm$ 0.13 \\
& & & B & 14.83 $\pm$ 0.32 \\
& & & g & 14.88 $\pm$ 0.13 \\
& & & r & 14.5 $\pm$ 0.13 \\
& & & i & 14.36 $\pm$ 0.13 \\
& & & z & 14.21 $\pm$ 0.13 \\
& & & J & 16.39 $\pm$ 0.19 \\
& & & H & 15.39 $\pm$ 0.13 \\
& & & K & 15.58 $\pm$ 0.13 \\ \hline
2006X & 185.7285 & 15.8218 & FUV & 14.12 $\pm$ 0.14 \\
& & & NUV & 13.22 $\pm$ 0.13 \\
& & & U & 10.69 $\pm$ 0.13 \\
& & & B & 9.78 $\pm$ 0.13 \\
& & & g & 10.74 $\pm$ 0.13 \\
& & & V & 9.28 $\pm$ 0.13 \\
& & & r & 10.27 $\pm$ 0.13 \\
& & & i & 12.44 $\pm$ 2.09 \\
& & & z & 9.75 $\pm$ 0.13 \\
& & & J & 9.36 $\pm$ 0.13 \\
& & & H & 9.2 $\pm$ 0.13 \\
& & & K & 9.38 $\pm$ 0.13 \\ \hline
1991bg & 186.2656 & 12.887 & FUV & 17.72 $\pm$ 0.13 \\
& & & NUV & 15.98 $\pm$ 0.13 \\
& & & U & 11.2 $\pm$ 0.13 \\
& & & B & 9.97 $\pm$ 0.13 \\
& & & g & 9.59 $\pm$ 0.13 \\
& & & V & 9.0 $\pm$ 0.13 \\
& & & r & 8.82 $\pm$ 0.13 \\
& & & I & 9.45 $\pm$ 0.13 \\
& & & z & 8.2 $\pm$ 0.13 \\
& & & J & 8.12 $\pm$ 0.13 \\
& & & H & 7.94 $\pm$ 0.13 \\
& & & K & 8.17 $\pm$ 0.13 \\ \hline
1999aa & 126.9303 & 21.5166 & FUV & 15.41 $\pm$ 0.13 \\
& & & NUV & 17.02 $\pm$ 0.13 \\
& & & u & 15.42 $\pm$ 0.13 \\
& & & U & 13.59 $\pm$ 0.2 \\
& & & B & 12.44 $\pm$ 0.21 \\
& & & g & 13.88 $\pm$ 0.13 \\
& & & V & 11.93 $\pm$ 0.21 \\
& & & r & 13.06 $\pm$ 0.13 \\
& & & i & 12.67 $\pm$ 0.13 \\
& & & z & 12.34 $\pm$ 0.13 \\
& & & J & 11.75 $\pm$ 0.13 \\
& & & H & 11.64 $\pm$ 0.13 \\
& & & K & 11.81 $\pm$ 0.13 \\
& & & W1 & 13.49 $\pm$ 0.13 \\
& & & W2 & 14.2 $\pm$ 0.13 \\
& & & W3 & 12.95 $\pm$ 0.13 \\
& & & W4 & 12.5 $\pm$ 0.13 \\ \hline
1998bu & 161.6906 & 11.8199 & FUV & 16.15 $\pm$ 0.13 \\
& & & NUV & 15.01 $\pm$ 0.13 \\
& & & U & 11.07 $\pm$ 0.17 \\
& & & B & 9.85 $\pm$ 0.17 \\
& & & V & 9.18 $\pm$ 0.17 \\
& & & r & 10.09 $\pm$ 0.13 \\
& & & i & 9.68 $\pm$ 0.13 \\
& & & z & 9.19 $\pm$ 0.13 \\
& & & J & 8.62 $\pm$ 0.13 \\
& & & H & 8.01 $\pm$ 0.13 \\
& & & K & 8.24 $\pm$ 0.13 \\ \hline
1998bp & 268.7115 & 18.3269 & B & 12.7 $\pm$ 0.25 \\
& & & V & 12.83 $\pm$ 0.14 \\
& & & J & 11.34 $\pm$ 0.13 \\
& & & H & 11.04 $\pm$ 0.13 \\
& & & K & 11.39 $\pm$ 0.13 \\
& & & W1 & 12.89 $\pm$ 0.13 \\
& & & W2 & 13.57 $\pm$ 0.13 \\ \hline
1998aq & 179.1172 & 55.1252 & u & 13.37 $\pm$ 0.13 \\
& & & B & 11.56 $\pm$ 0.23 \\
& & & g & 12.16 $\pm$ 0.13 \\
& & & V & 11.55 $\pm$ 0.13 \\
& & & r & 11.56 $\pm$ 0.13 \\
& & & i & 11.28 $\pm$ 0.13 \\
& & & z & 11.11 $\pm$ 0.13 \\
& & & J & 10.73 $\pm$ 0.13 \\
& & & H & 10.54 $\pm$ 0.13 \\
& & & K & 10.76 $\pm$ 0.13 \\
& & & W1 & 13.01 $\pm$ 0.13 \\
& & & W2 & 13.38 $\pm$ 0.13 \\
& & & W3 & 11.23 $\pm$ 0.13 \\
& & & W4 & 9.72 $\pm$ 0.13 \\ \hline
1996X & 199.5211 & -26.8372 & V & 10.17 $\pm$ 0.17 \\
& & & r & 10.73 $\pm$ 0.13 \\
& & & J & 9.13 $\pm$ 0.13 \\
& & & H & 8.96 $\pm$ 0.13 \\
& & & K & 9.19 $\pm$ 0.13 \\
& & & W1 & 11.32 $\pm$ 0.13 \\
& & & W2 & 11.97 $\pm$ 0.13 \\
& & & W3 & 13.06 $\pm$ 0.13 \\
& & & W4 & 13.2 $\pm$ 0.13 \\ \hline
1995D & 145.221 & 5.1658 & B & 12.58 $\pm$ 0.17 \\
& & & g & 12.8 $\pm$ 0.13 \\
& & & V & 11.77 $\pm$ 0.17 \\
& & & i & 11.51 $\pm$ 0.13 \\
& & & z & 11.2 $\pm$ 0.13 \\
& & & J & 10.64 $\pm$ 0.13 \\
& & & H & 10.42 $\pm$ 0.13 \\
& & & K & 10.62 $\pm$ 0.13 \\
& & & W1 & 12.56 $\pm$ 0.13 \\
& & & W2 & 13.24 $\pm$ 0.13 \\
& & & W3 & 14.07 $\pm$ 0.13 \\
& & & W4 & 13.85 $\pm$ 0.15 \\ \hline
1994ae & 161.7669 & 17.2736 & FUV & 14.27 $\pm$ 0.13 \\
& & & NUV & 15.6 $\pm$ 0.13 \\
& & & u & 13.92 $\pm$ 0.13 \\
& & & B & 11.99 $\pm$ 0.21 \\
& & & g & 12.68 $\pm$ 0.13 \\
& & & r & 12.07 $\pm$ 0.13 \\
& & & i & 11.74 $\pm$ 0.13 \\
& & & I & 11.46 $\pm$ 0.13 \\
& & & z & 11.48 $\pm$ 0.13 \\
& & & J & 11.34 $\pm$ 0.13 \\
& & & H & 11.16 $\pm$ 0.13 \\
& & & K & 11.26 $\pm$ 0.13 \\
& & & W1 & 13.33 $\pm$ 0.13 \\
& & & W2 & 13.87 $\pm$ 0.13 \\
& & & W3 & 12.06 $\pm$ 0.13 \\
& & & W4 & 11.31 $\pm$ 0.13 \\ \hline
1990N & 190.7183 & 13.2574 & FUV & 14.37 $\pm$ 0.13 \\
& & & NUV & 14.05 $\pm$ 0.13 \\
& & & u & 12.88 $\pm$ 0.13 \\
& & & U & 12.96 $\pm$ 0.13 \\
& & & B & 11.97 $\pm$ 0.13 \\
& & & g & 11.66 $\pm$ 0.13 \\
& & & V & 11.47 $\pm$ 0.13 \\
& & & I & 10.78 $\pm$ 0.13 \\
& & & J & 10.65 $\pm$ 0.13 \\
& & & H & 10.51 $\pm$ 0.13 \\
& & & K & 10.75 $\pm$ 0.13 \\
& & & W1 & 12.76 $\pm$ 0.34 \\
& & & W2 & 13.32 $\pm$ 0.13 \\
& & & W3 & 13.52 $\pm$ 0.13 \\
& & & W4 & 12.99 $\pm$ 0.13 \\ \hline
1986G & 201.3651 & -43.0191 & FUV & 13.36 $\pm$ 0.13 \\
& & & NUV & 11.69 $\pm$ 0.13 \\
& & & B & 7.25 $\pm$ 0.13 \\
& & & V & 6.49 $\pm$ 0.13 \\
& & & R & 6.22 $\pm$ 0.13 \\
& & & J & 5.84 $\pm$ 0.13 \\
& & & H & 5.63 $\pm$ 0.13 \\
& & & K & 5.83 $\pm$ 0.13 \\
& & & W1 & 8.39 $\pm$ 0.13 \\
& & & W2 & 8.5 $\pm$ 0.13 \\
& & & W3 & 7.78 $\pm$ 0.13 \\
& & & W4 & 6.87 $\pm$ 0.13 \\ \hline
2011fe & 210.8023 & 54.349 & FUV & 12.19 $\pm$ 0.13 \\
& & & NUV & 11.51 $\pm$ 0.13 \\
& & & B & 3.11 $\pm$ 0.13 \\
& & & V & 7.83 $\pm$ 0.13 \\
& & & R & 7.74 $\pm$ 0.13 \\
& & & J & 7.46 $\pm$ 0.13 \\
& & & H & 7.23 $\pm$ 0.13 \\
& & & K & 7.42 $\pm$ 0.13 \\
& & & W1 & 7.88 $\pm$ 0.13 \\
& & & W2 & 8.36 $\pm$ 0.13 \\
& & & W3 & 6.59 $\pm$ 0.13 \\
& & & W4 & 6.61 $\pm$ 0.13 \\ \hline
2008Q & 21.1988 & 9.5388 & FUV & 18.03 $\pm$ 0.13 \\
& & & NUV & 17.11 $\pm$ 0.13 \\
& & & V & 10.02 $\pm$ 0.13 \\
& & & R & 9.99 $\pm$ 0.13 \\
& & & J & 9.03 $\pm$ 0.13 \\
& & & H & 8.83 $\pm$ 0.13 \\
& & & K & 9.05 $\pm$ 0.13 \\
& & & W1 & 11.45 $\pm$ 0.13 \\
& & & W2 & 12.09 $\pm$ 0.13 \\
& & & W3 & 12.49 $\pm$ 0.13 \\
& & & W4 & 12.7 $\pm$ 0.13 \\ \hline
2006D & 193.1446 & 9.7767 & FUV & 16.09 $\pm$ 0.13 \\
& & & NUV & 15.7 $\pm$ 0.13 \\
& & & J & 12.53 $\pm$ 0.13 \\
& & & H & 12.34 $\pm$ 0.13 \\
& & & K & 12.48 $\pm$ 0.13 \\
& & & W1 & 13.24 $\pm$ 0.13 \\
& & & W2 & 13.64 $\pm$ 0.13 \\
& & & W3 & 11.16 $\pm$ 0.13 \\
& & & W4 & 9.55 $\pm$ 0.13 \\ \hline
2005ke & 53.7556 & -24.9332 & FUV & 14.27 $\pm$ 0.13 \\
& & & NUV & 13.94 $\pm$ 0.13 \\
& & & U & 12.69 $\pm$ 0.14 \\
& & & B & 11.31 $\pm$ 0.14 \\
& & & V & 10.6 $\pm$ 0.14 \\
& & & R & 10.1 $\pm$ 0.13 \\
& & & J & 9.5 $\pm$ 0.13 \\
& & & H & 9.32 $\pm$ 0.13 \\
& & & K & 9.52 $\pm$ 0.13 \\
& & & W1 & 12.34 $\pm$ 0.13 \\
& & & W2 & 12.93 $\pm$ 0.13 \\
& & & W3 & 13.68 $\pm$ 0.13 \\
& & & W4 & 13.24 $\pm$ 0.13 \\ \hline
2003hv & 46.0332 & -26.0696 & U & 12.88 $\pm$ 0.19 \\
& & & B & 11.44 $\pm$ 0.19 \\
& & & V & 10.68 $\pm$ 0.19 \\
& & & R & 10.5 $\pm$ 0.13 \\
& & & J & 9.53 $\pm$ 0.13 \\
& & & H & 9.33 $\pm$ 0.13 \\
& & & K & 9.57 $\pm$ 0.13 \\
& & & W1 & 11.53 $\pm$ 0.13 \\
& & & W2 & 12.2 $\pm$ 0.13 \\
& & & W3 & 13.5 $\pm$ 0.13 \\
& & & W4 & 13.68 $\pm$ 0.13 \\ \hline
2003du & 218.6542 & 59.3378 & u & 16.16 $\pm$ 0.13 \\
& & & B & 14.58 $\pm$ 0.21 \\
& & & g & 14.86 $\pm$ 0.13 \\
& & & V & 14.24 $\pm$ 0.22 \\
& & & r & 14.51 $\pm$ 0.13 \\
& & & i & 14.55 $\pm$ 0.13 \\
& & & z & 14.32 $\pm$ 0.13 \\ \hline
2002dp & 352.1162 & 22.4212 & FUV & 14.68 $\pm$ 0.11 \\
& & & NUV & 14.25 $\pm$ 0.35 \\
& & & B & 12.06 $\pm$ 0.23 \\
& & & V & 11.69 $\pm$ 0.25 \\
& & & J & 11.17 $\pm$ 0.13 \\
& & & H & 11.01 $\pm$ 0.13 \\
& & & K & 11.16 $\pm$ 0.13 \\
& & & W1 & 13.36 $\pm$ 0.13 \\
& & & W2 & 13.78 $\pm$ 0.13 \\
& & & W3 & 11.44 $\pm$ 0.13 \\
& & & W4 & 9.97 $\pm$ 0.13 \\ \hline
2002cs & 281.7399 & 45.7057 & NUV & 18.95 $\pm$ 0.13 \\
& & & U & 14.03 $\pm$ 0.22 \\
& & & B & 12.67 $\pm$ 0.21 \\
& & & V & 11.94 $\pm$ 0.21 \\
& & & r & 13.24 $\pm$ 0.13 \\
& & & J & 11.1 $\pm$ 0.13 \\
& & & H & 10.86 $\pm$ 0.13 \\
& & & K & 11.11 $\pm$ 0.13 \\
& & & W1 & 12.77 $\pm$ 0.13 \\
& & & W2 & 13.44 $\pm$ 0.13 \\
& & & W3 & 14.52 $\pm$ 0.13 \\
& & & W4 & 14.66 $\pm$ 0.21 \\ \hline
2000E & 309.3087 & 65.1056 & FUV & 14.47 $\pm$ 0.13 \\
& & & NUV & 13.39 $\pm$ 0.13 \\
& & & B & 9.82 $\pm$ 0.21 \\
& & & V & 9.4 $\pm$ 0.22 \\
& & & J & 9.01 $\pm$ 0.13 \\
& & & H & 8.8 $\pm$ 0.13 \\
& & & K & 9.06 $\pm$ 0.13 \\
& & & W1 & 11.65 $\pm$ 0.13 \\
& & & W2 & 12.1 $\pm$ 0.13 \\
& & & W3 & 10.32 $\pm$ 0.13 \\
& & & W4 & 9.05 $\pm$ 0.13 \\ \hline
1999by & 140.511 & 50.9765 & FUV & 15.9 $\pm$ 0.13 \\
& & & NUV & 14.94 $\pm$ 0.13 \\
& & & U & 11.12 $\pm$ 0.13 \\
& & & B & 9.86 $\pm$ 0.13 \\
& & & g & 9.74 $\pm$ 0.9 \\
& & & V & 9.16 $\pm$ 0.13 \\
& & & r & 9.0 $\pm$ 0.75 \\
& & & I & 8.41 $\pm$ 0.13 \\
& & & J & 7.97 $\pm$ 0.13 \\
& & & H & 7.74 $\pm$ 0.13 \\
& & & K & 7.98 $\pm$ 0.13 \\
& & & W1 & 10.95 $\pm$ 0.13 \\
& & & W2 & 11.63 $\pm$ 0.13 \\
& & & W3 & 12.69 $\pm$ 0.13 \\
& & & W4 & 12.74 $\pm$ 0.13 \\ \hline
1991T & 188.5351 & 2.6537 & FUV & 14.95 $\pm$ 0.13 \\
& & & NUV & 14.27 $\pm$ 0.13 \\
& & & u & 12.56 $\pm$ 0.13 \\
& & & U & 12.25 $\pm$ 0.17 \\
& & & B & 11.12 $\pm$ 0.13 \\
& & & g & 10.89 $\pm$ 0.13 \\
& & & V & 10.45 $\pm$ 0.13 \\
& & & r & 10.14 $\pm$ 0.13 \\
& & & i & 9.69 $\pm$ 0.13 \\
& & & z & 9.36 $\pm$ 0.13 \\
& & & J & 9.01 $\pm$ 0.13 \\
& & & H & 8.71 $\pm$ 0.13 \\
& & & K & 8.86 $\pm$ 0.13 \\ \hline
\end{longtable}
}
\vspace{-0.1in}
\tablefoot{The SNe names and the respective coordinates (R.A. and Decl.), filters, and photometric magnitudes of their host galaxies. All photometry is from NED and corrected for Galactic extinction in the direction in the direction of the SNe.}

\section{Host Galaxy Properties} \label{appendix}
Here, we record all stellar population properties of the host galaxies of SNe~Ia studied in this work.

\footnotesize{
\begin{longtable}{l|ccccccc}
\caption{\label{tab:hostdata} Host Data}\\
\hline \hline 
SN Name & 
Type &
$z$ & 
$t_m$ &
log(M$_*$/M$_\odot$) &
SFR [M$_\odot$/yr] & 
log(Z$_*$/Z$_\odot$) & 
A$_V$ [mag] \\
\hline 
\endfirsthead
\caption{continued.} \\
\hline \hline
SN Name & 
Type &
$z$ & 
$t_m$ &
log(M$_*$/M$_\odot$) &
SFR [M$_\odot$/yr] & 
log(Z$_*$/Z$_\odot$) & 
A$_V$ [mag] \\
\hline
\endhead
\hline
\endfoot
1998dh & low & 0.009 & $3.25^{+1.97}_{-1.84}$ & $10.52^{+0.15}_{-0.21}$ & $9.11^{+7.05}_{-6.37}$ & $-0.0^{+0.12}_{-0.13}$ & $1.63^{+0.69}_{-0.72}$ \\
1998dm & low & 0.0065 & $0.72^{+0.02}_{-0.03}$ & $8.93^{+0.01}_{-0.01}$ & $0.02^{+0.01}_{-0.0}$ & $-1.19^{+0.01}_{-0.01}$ & $0.09^{+0.03}_{-0.03}$ \\
1999cp & low & 0.0095 & $0.54^{+0.23}_{-0.13}$ & $9.31^{+0.1}_{-0.07}$ & $4.3^{+0.32}_{-0.4}$ & $-0.54^{+0.14}_{-0.22}$ & $0.33^{+0.12}_{-0.1}$ \\
1999dq & low & 0.0143 & $4.72^{+1.67}_{-1.15}$ & $10.68^{+0.06}_{-0.06}$ & $4.24^{+0.56}_{-0.66}$ & $-0.34^{+0.1}_{-0.11}$ & $1.44^{+0.17}_{-0.19}$ \\
1999gp & low & 0.0267 & $3.05^{+2.02}_{-1.24}$ & $10.47^{+0.11}_{-0.1}$ & $2.55^{+1.89}_{-1.57}$ & $-0.26^{+0.1}_{-0.12}$ & $0.16^{+0.17}_{-0.07}$ \\
2000cx & low & 0.008 & $6.87^{+0.75}_{-0.72}$ & $10.86^{+0.03}_{-0.03}$ & $0.01^{+0.0}_{-0.0}$ & $-0.49^{+0.05}_{-0.05}$ & $0.08^{+0.03}_{-0.02}$ \\
2000dn & low & 0.0321 & $2.98^{+0.98}_{-0.62}$ & $10.56^{+0.08}_{-0.07}$ & $0.0^{+0.06}_{-0.0}$ & $-0.58^{+0.09}_{-0.11}$ & $0.21^{+0.14}_{-0.08}$ \\
2000dr & low & 0.0188 & $8.19^{+1.29}_{-1.22}$ & $10.66^{+0.04}_{-0.05}$ & $0.02^{+0.01}_{-0.01}$ & $-0.44^{+0.08}_{-0.07}$ & $0.11^{+0.06}_{-0.04}$ \\
2000fa & low & 0.0213 & $0.03^{+0.03}_{-0.02}$ & $9.27^{+0.17}_{-0.17}$ & $57.27^{+16.08}_{-14.02}$ & $-1.01^{+0.17}_{-0.12}$ & $3.12^{+0.59}_{-0.48}$ \\
2001en & high & 0.0159 & $4.61^{+1.33}_{-1.0}$ & $10.09^{+0.05}_{-0.05}$ & $1.04^{+0.12}_{-0.11}$ & $-0.96^{+0.12}_{-0.09}$ & $1.77^{+0.2}_{-0.21}$ \\
2001ep & low & 0.0131 & $0.61^{+0.06}_{-0.05}$ & $9.49^{+0.02}_{-0.02}$ & $0.23^{+0.11}_{-0.07}$ & $-1.17^{+0.04}_{-0.02}$ & $0.35^{+0.12}_{-0.09}$ \\
2002bo & high & 0.0044 & $7.3^{+1.41}_{-0.88}$ & $10.43^{+0.05}_{-0.04}$ & $0.12^{+0.02}_{-0.02}$ & $-0.76^{+0.06}_{-0.06}$ & $0.47^{+0.07}_{-0.07}$ \\
2002cr & low & 0.0095 & $0.19^{+0.02}_{-0.02}$ & $9.06^{+0.05}_{-0.05}$ & $2.62^{+0.19}_{-0.19}$ & $-1.04^{+0.07}_{-0.1}$ & $0.18^{+0.04}_{-0.03}$ \\
2002dj & high & 0.0094 & $3.28^{+0.09}_{-0.18}$ & $10.9^{+0.01}_{-0.03}$ & $0.0^{+0.15}_{-0.0}$ & $-1.0^{+0.0}_{-0.0}$ & $0.19^{+0.03}_{-0.03}$ \\
2002dl & low & 0.0163 & $1.56^{+0.75}_{-0.33}$ & $9.91^{+0.09}_{-0.06}$ & $0.26^{+0.16}_{-0.13}$ & $-1.17^{+0.05}_{-0.02}$ & $0.69^{+0.16}_{-0.14}$ \\
2002eb & low & 0.0275 & $0.84^{+0.16}_{-0.17}$ & $9.65^{+0.04}_{-0.05}$ & $6.04^{+0.35}_{-0.33}$ & $-1.1^{+0.1}_{-0.06}$ & $2.6^{+0.28}_{-0.23}$ \\
2002er & low & 0.0085 & $0.75^{+0.26}_{-0.32}$ & $9.33^{+0.05}_{-0.07}$ & $2.79^{+0.22}_{-0.29}$ & $-1.02^{+0.15}_{-0.11}$ & $2.57^{+0.33}_{-0.3}$ \\
2002fk & low & 0.0071 & $0.69^{+0.02}_{-0.02}$ & $9.39^{+0.01}_{-0.01}$ & $0.08^{+0.01}_{-0.01}$ & $-1.18^{+0.03}_{-0.01}$ & $0.3^{+0.04}_{-0.04}$ \\
2002ha & low & 0.014 & $9.86^{+0.56}_{-1.0}$ & $10.8^{+0.02}_{-0.04}$ & $0.08^{+0.02}_{-0.01}$ & $-0.84^{+0.06}_{-0.07}$ & $0.07^{+0.03}_{-0.03}$ \\
2002he & high & 0.0246 & $12.12^{+0.6}_{-0.84}$ & $10.89^{+0.02}_{-0.03}$ & $0.0^{+0.0}_{-0.0}$ & $0.02^{+0.06}_{-0.06}$ & $0.01^{+0.03}_{-0.01}$ \\
2003cg & low & 0.0041 & $0.74^{+0.06}_{-0.05}$ & $9.83^{+0.02}_{-0.02}$ & $0.32^{+0.06}_{-0.05}$ & $0.26^{+0.03}_{-0.05}$ & $2.32^{+0.27}_{-0.29}$ \\
2003fa & low & 0.0404 & $1.02^{+0.3}_{-0.32}$ & $10.08^{+0.06}_{-0.06}$ & $13.35^{+1.19}_{-1.16}$ & $-0.56^{+0.1}_{-0.12}$ & $1.2^{+0.32}_{-0.27}$ \\
2003gn & low & 0.0344 & $1.89^{+0.5}_{-0.51}$ & $9.49^{+0.06}_{-0.06}$ & $1.9^{+0.22}_{-0.13}$ & $0.24^{+0.04}_{-0.07}$ & $1.01^{+0.12}_{-0.08}$ \\
2003gt & low & 0.0152 & $1.66^{+1.73}_{-0.86}$ & $9.67^{+0.11}_{-0.06}$ & $1.49^{+0.34}_{-0.42}$ & $-0.28^{+0.27}_{-0.28}$ & $2.39^{+0.57}_{-0.39}$ \\
2004at & low & 0.0224 & $0.59^{+0.05}_{-0.06}$ & $8.99^{+0.02}_{-0.03}$ & $0.09^{+0.06}_{-0.03}$ & $-1.16^{+0.05}_{-0.03}$ & $0.1^{+0.12}_{-0.07}$ \\
2004dt & high & 0.0198 & $9.7^{+0.47}_{-0.9}$ & $10.61^{+0.02}_{-0.03}$ & $0.1^{+0.03}_{-0.03}$ & $-0.98^{+0.06}_{-0.04}$ & $0.29^{+0.07}_{-0.07}$ \\
2004ef & high & 0.031 & $1.43^{+0.33}_{-0.27}$ & $9.66^{+0.04}_{-0.02}$ & $0.25^{+0.06}_{-0.05}$ & $0.26^{+0.02}_{-0.04}$ & $0.14^{+0.12}_{-0.09}$ \\
2004eo & low & 0.0156 & $2.19^{+1.2}_{-0.54}$ & $10.56^{+0.09}_{-0.06}$ & $1.16^{+0.64}_{-0.58}$ & $-0.75^{+0.12}_{-0.13}$ & $0.53^{+0.11}_{-0.13}$ \\
2005cf & low & 0.0064 & $0.26^{+0.03}_{-0.03}$ & $8.76^{+0.05}_{-0.07}$ & $0.76^{+0.09}_{-0.07}$ & $-1.05^{+0.12}_{-0.1}$ & $0.54^{+0.1}_{-0.1}$ \\
2005de & low & 0.0153 & $4.08^{+3.01}_{-2.36}$ & $10.19^{+0.22}_{-0.21}$ & $2.08^{+3.32}_{-2.08}$ & $-0.04^{+0.2}_{-0.24}$ & $1.36^{+0.92}_{-0.89}$ \\
2005ki & low & 0.0195 & $2.18^{+0.8}_{-0.53}$ & $10.57^{+0.08}_{-0.08}$ & $0.0^{+0.02}_{-0.0}$ & $-0.99^{+0.11}_{-0.12}$ & $0.85^{+0.21}_{-0.14}$ \\
2006cp & high & 0.0223 & $1.95^{+1.88}_{-1.1}$ & $10.07^{+0.15}_{-0.15}$ & $6.28^{+5.29}_{-2.29}$ & $-0.03^{+0.22}_{-0.35}$ & $1.18^{+0.72}_{-0.5}$ \\
2006gr & low & 0.0347 & $1.22^{+0.8}_{-0.57}$ & $10.05^{+0.07}_{-0.06}$ & $7.66^{+1.12}_{-1.44}$ & $0.07^{+0.12}_{-0.17}$ & $1.8^{+0.42}_{-0.35}$ \\
2006le & low & 0.0174 & $0.57^{+0.06}_{-0.05}$ & $10.25^{+0.03}_{-0.03}$ & $1.89^{+0.72}_{-0.51}$ & $-1.07^{+0.09}_{-0.09}$ & $0.4^{+0.16}_{-0.11}$ \\
2007af & low & 0.0055 & $0.01^{+0.0}_{-0.0}$ & $9.36^{+0.09}_{-0.7}$ & $296.92^{+68.29}_{-265.39}$ & $-1.06^{+0.8}_{-0.07}$ & $6.0^{+0.29}_{-2.04}$ \\
2007le & high & 0.0067 & $3.34^{+1.56}_{-1.42}$ & $9.57^{+0.07}_{-0.1}$ & $0.64^{+0.06}_{-0.06}$ & $-1.14^{+0.08}_{-0.05}$ & $0.87^{+0.12}_{-0.11}$ \\
2007qe & high & 0.024 & $0.13^{+0.05}_{-0.02}$ & $8.13^{+0.06}_{-0.06}$ & $0.91^{+0.09}_{-0.09}$ & $-1.05^{+0.05}_{-0.08}$ & $1.23^{+0.12}_{-0.12}$ \\
2008bf & low & 0.0244 & $0.95^{+0.06}_{-0.05}$ & $10.34^{+0.01}_{-0.02}$ & $0.18^{+0.23}_{-0.09}$ & $-0.12^{+0.1}_{-0.11}$ & $0.44^{+0.11}_{-0.1}$ \\
2008ec & low & 0.0163 & $0.9^{+0.78}_{-0.39}$ & $10.44^{+0.12}_{-0.11}$ & $35.13^{+14.03}_{-11.54}$ & $0.23^{+0.05}_{-0.08}$ & $2.48^{+0.47}_{-0.37}$ \\
2001V & low & 0.015 & $1.26^{+1.35}_{-0.54}$ & $8.99^{+0.14}_{-0.05}$ & $0.13^{+0.11}_{-0.05}$ & $-1.16^{+0.05}_{-0.03}$ & $0.39^{+0.23}_{-0.15}$ \\
2005hk & low & 0.013 & $0.42^{+0.02}_{-0.02}$ & $8.53^{+0.02}_{-0.02}$ & $0.1^{+0.01}_{-0.01}$ & $-1.19^{+0.01}_{-0.01}$ & $0.07^{+0.02}_{-0.02}$ \\
2006ax & low & 0.0167 & $6.11^{+2.74}_{-2.21}$ & $9.89^{+0.1}_{-0.13}$ & $0.0^{+0.01}_{-0.0}$ & $-0.54^{+0.22}_{-0.18}$ & $1.16^{+0.35}_{-0.41}$ \\
2006lf & low & 0.0132 & $0.52^{+0.04}_{-0.03}$ & $9.87^{+0.02}_{-0.02}$ & $0.94^{+0.26}_{-0.2}$ & $-1.18^{+0.02}_{-0.01}$ & $0.2^{+0.06}_{-0.05}$ \\
2007bd & high & 0.0257 & $1.29^{+0.29}_{-0.49}$ & $10.09^{+0.05}_{-0.04}$ & $0.1^{+0.24}_{-0.09}$ & $-1.16^{+0.05}_{-0.03}$ & $0.07^{+0.07}_{-0.04}$ \\
2007ci & low & 0.018 & $8.61^{+1.45}_{-1.2}$ & $10.69^{+0.04}_{-0.04}$ & $0.01^{+0.01}_{-0.01}$ & $-0.36^{+0.05}_{-0.06}$ & $0.01^{+0.01}_{-0.0}$ \\
2005kc & low & 0.0151 & $3.12^{+1.92}_{-1.72}$ & $10.64^{+0.17}_{-0.16}$ & $12.27^{+8.44}_{-7.79}$ & $0.04^{+0.14}_{-0.17}$ & $1.73^{+0.46}_{-0.52}$ \\
2007on & low & 0.0065 & $10.32^{+0.48}_{-0.78}$ & $10.83^{+0.02}_{-0.03}$ & $0.04^{+0.0}_{-0.0}$ & $-0.47^{+0.05}_{-0.04}$ & $0.04^{+0.01}_{-0.01}$ \\
2008gp & low & 0.033 & $0.64^{+0.28}_{-0.13}$ & $9.88^{+0.04}_{-0.05}$ & $2.61^{+1.12}_{-0.75}$ & $0.13^{+0.1}_{-0.15}$ & $1.53^{+0.42}_{-0.37}$ \\
2008hv & low & 0.0126 & $3.47^{+0.87}_{-0.62}$ & $10.27^{+0.06}_{-0.06}$ & $0.0^{+0.0}_{-0.0}$ & $-0.93^{+0.09}_{-0.07}$ & $0.02^{+0.02}_{-0.01}$ \\
2003W & high & 0.0201 & $2.03^{+1.0}_{-0.79}$ & $10.29^{+0.1}_{-0.11}$ & $10.22^{+3.2}_{-2.83}$ & $0.04^{+0.11}_{-0.18}$ & $0.99^{+0.28}_{-0.18}$ \\
2003Y & faint & 0.0169 & $5.58^{+0.94}_{-1.2}$ & $10.52^{+0.05}_{-0.07}$ & $0.0^{+0.0}_{-0.0}$ & $-0.74^{+0.08}_{-0.07}$ & $0.05^{+0.04}_{-0.03}$ \\
2005M & low & 0.0246 & $0.28^{+0.08}_{-0.11}$ & $8.84^{+0.1}_{-0.13}$ & $2.05^{+1.04}_{-0.94}$ & $-1.09^{+0.11}_{-0.08}$ & $0.2^{+0.17}_{-0.17}$ \\
2006X & high & 0.0052 & $0.36^{+0.16}_{-0.12}$ & $10.11^{+0.28}_{-0.12}$ & $24.81^{+7.75}_{-17.63}$ & $0.26^{+0.03}_{-0.06}$ & $1.68^{+0.28}_{-0.23}$ \\
1991bg & faint & 0.0034 & $5.8^{+0.41}_{-0.37}$ & $10.75^{+0.02}_{-0.02}$ & $0.0^{+0.0}_{-0.0}$ & $0.28^{+0.02}_{-0.02}$ & $0.02^{+0.02}_{-0.01}$ \\
1999aa & low & 0.0145 & $5.85^{+0.97}_{-1.05}$ & $10.28^{+0.04}_{-0.05}$ & $0.73^{+0.08}_{-0.07}$ & $-0.92^{+0.09}_{-0.07}$ & $0.88^{+0.16}_{-0.17}$ \\
1998bu & low & 0.003 & $1.87^{+1.43}_{-0.47}$ & $10.35^{+0.09}_{-0.06}$ & $3.79^{+3.38}_{-2.4}$ & $0.1^{+0.13}_{-0.11}$ & $3.21^{+0.61}_{-0.63}$ \\
1998bp & low & 0.0104 & $4.07^{+1.41}_{-1.1}$ & $10.19^{+0.08}_{-0.09}$ & $0.0^{+0.01}_{-0.0}$ & $-0.83^{+0.1}_{-0.12}$ & $0.08^{+0.17}_{-0.06}$ \\
1998aq & low & 0.0037 & $0.42^{+0.02}_{-0.02}$ & $8.98^{+0.02}_{-0.02}$ & $0.32^{+0.03}_{-0.03}$ & $-1.19^{+0.01}_{-0.01}$ & $0.99^{+0.1}_{-0.14}$ \\
1996X & low & 0.0069 & $2.52^{+0.62}_{-0.54}$ & $10.34^{+0.07}_{-0.07}$ & $0.0^{+0.02}_{-0.0}$ & $-0.78^{+0.06}_{-0.07}$ & $0.08^{+0.02}_{-0.02}$ \\
1995D & low & 0.0066 & $8.08^{+1.46}_{-1.15}$ & $10.14^{+0.05}_{-0.05}$ & $0.0^{+0.0}_{-0.0}$ & $-1.01^{+0.06}_{-0.07}$ & $0.2^{+0.04}_{-0.04}$ \\
1994ae & low & 0.0043 & $0.6^{+0.03}_{-0.03}$ & $9.03^{+0.01}_{-0.01}$ & $0.08^{+0.02}_{-0.01}$ & $-1.18^{+0.02}_{-0.01}$ & $0.82^{+0.14}_{-0.12}$ \\
1990N & low & 0.0034 & $2.39^{+0.61}_{-0.67}$ & $9.29^{+0.06}_{-0.08}$ & $0.08^{+0.01}_{-0.02}$ & $-1.17^{+0.03}_{-0.02}$ & $0.07^{+0.02}_{-0.01}$ \\
1986G & low & 0.0018 & $0.76^{+0.01}_{-0.01}$ & $10.31^{+0.01}_{-0.01}$ & $0.29^{+0.04}_{-0.03}$ & $-1.18^{+0.02}_{-0.01}$ & $0.91^{+0.08}_{-0.16}$ \\
2011fe & low & 0.0008 & $0.79^{+0.09}_{-0.09}$ & $9.42^{+0.02}_{-0.02}$ & $0.33^{+0.06}_{-0.06}$ & $0.01^{+0.12}_{-0.13}$ & $0.82^{+0.16}_{-0.13}$ \\
2008q & low & 0.008 & $10.62^{+0.21}_{-0.41}$ & $10.77^{+0.01}_{-0.02}$ & $0.05^{+0.01}_{-0.01}$ & $-0.34^{+0.05}_{-0.05}$ & $0.19^{+0.05}_{-0.05}$ \\
2006D & low & 0.0085 & $0.55^{+0.2}_{-0.21}$ & $9.26^{+0.08}_{-0.11}$ & $3.69^{+0.34}_{-0.29}$ & $-0.91^{+0.21}_{-0.16}$ & $2.33^{+0.24}_{-0.25}$ \\
2005ke & faint & 0.0049 & $1.08^{+0.31}_{-0.19}$ & $9.47^{+0.06}_{-0.03}$ & $0.05^{+0.04}_{-0.03}$ & $-1.17^{+0.05}_{-0.02}$ & $0.07^{+0.02}_{-0.01}$ \\
2003hv & low & 0.0056 & $3.56^{+0.78}_{-0.59}$ & $10.21^{+0.05}_{-0.06}$ & $0.0^{+0.0}_{-0.0}$ & $-1.03^{+0.05}_{-0.07}$ & $0.05^{+0.01}_{-0.01}$ \\
2003du & low & 0.0064 & $0.53^{+0.56}_{-0.29}$ & $8.2^{+0.15}_{-0.16}$ & $0.33^{+0.18}_{-0.13}$ & $-0.79^{+0.2}_{-0.2}$ & $1.14^{+0.31}_{-0.3}$ \\
2002dp & low & 0.0116 & $0.28^{+0.03}_{-0.02}$ & $9.62^{+0.04}_{-0.05}$ & $4.41^{+0.4}_{-0.33}$ & $-1.11^{+0.09}_{-0.06}$ & $0.49^{+0.08}_{-0.06}$ \\
2002cs & high & 0.0158 & $10.52^{+0.34}_{-0.75}$ & $10.74^{+0.02}_{-0.02}$ & $0.02^{+0.01}_{-0.01}$ & $-0.07^{+0.06}_{-0.07}$ & $0.11^{+0.06}_{-0.05}$ \\
2000E & high & 0.0048 & $0.46^{+0.05}_{-0.03}$ & $9.68^{+0.02}_{-0.03}$ & $1.36^{+0.18}_{-0.15}$ & $-1.15^{+0.06}_{-0.03}$ & $0.77^{+0.09}_{-0.09}$ \\
1999by & faint & 0.0021 & $10.63^{+0.17}_{-0.33}$ & $9.9^{+0.01}_{-0.01}$ & $0.01^{+0.0}_{-0.0}$ & $-1.11^{+0.04}_{-0.04}$ & $0.08^{+0.01}_{-0.01}$ \\
1991T & low & 0.0058 & $5.84^{+1.74}_{-1.78}$ & $10.92^{+0.05}_{-0.07}$ & $2.24^{+1.22}_{-0.7}$ & $0.06^{+0.15}_{-0.14}$ & $1.02^{+0.65}_{-0.4}$ \\ \hline
\end{longtable}
}
\vspace{-0.1in}
\tablefoot{The types of each SNe~Ia studied in this work (low velocity and high velocity and or faint) the host galaxy redshifts, stellar population ages ($t_m$), stellar masses (log(M$_*$/M$_\odot$), SFRs, stellar metallicities (log(Z$_*$/Z$_\odot$), and total dust extinction ($A_V$). All redshifts are from NED.}

\end{appendix}
\end{document}